\documentclass[superscriptaddress,nofootinbib,tightenlines,preprint,12pt]{article}
\pdfoutput=1
\usepackage[utf8]{inputenc}
\usepackage[left=1in, right=1in, top=1in, bottom=1in]{geometry}

\usepackage[pdftex,breaklinks,colorlinks,linktocpage,
citecolor=blue,
urlcolor=blue]{hyperref}

\usepackage{bm}
\usepackage{amssymb}
\usepackage{amsfonts}
\usepackage{mathrsfs}
\usepackage{amsmath}
\usepackage{amsthm}
\usepackage{nicefrac}
\usepackage{xcolor}
\usepackage{braket}
\usepackage{graphicx}
\usepackage{subcaption}
\usepackage{comment}
\usepackage{authblk}

\newcommand{\be}{\begin{equation}}
\newcommand{\ee}{\end{equation}}
\newcommand{\bea}{\begin{eqnarray}}
\newcommand{\eea}{\end{eqnarray}}
\newcommand{\munu}{{\mu\nu}}
\DeclareMathOperator{\tr}{tr}
\def\bml{\begin{subequations}}
\def\blea{\bml\begin{eqnarray}}
\def\eml{\end{subequations}}
\def\elea{\end{eqnarray}\eml}

\theoremstyle{definition}
\newtheorem{definition}{Definition}[section]

  \newcommand{\dif}{\mathop{}\!\mathrm{d}}

\date{\today}

\begin{document}

\title{A new approach in classical Klein-Gordon cosmology: ``Small Bangs'', inflation and Dark Energy}

\author[1]{Eleni-Alexandra Kontou \thanks{\href{eleni.kontou@kcl.ac.uk}{eleni.kontou@kcl.ac.uk}}}
\author[2]{Nicolai Rothe \thanks{\href{rothe@math.tu-berlin.de}{rothe@math.tu-berlin.de}}}
\affil[1]{Department of Mathematics, King’s College London, Strand, London WC2R 2LS, United Kingdom}
\affil[2]{Institute of Mathematics, Technische Universität Berlin, 10623 Berlin, Germany}

\maketitle

\begin{abstract}

In this work, we analyze the cosmological model in which the expansion is driven by a classical, free Klein-Gordon field on a flat, four-dimensional Friedmann–Lemaître– Robertson–Walker spacetime. The model allows for arbitrary mass, non-zero cosmological constant and coupling to curvature. We find that there are strong restrictions to the parameter space, due to the requirement for the reality of the field values. At early cosmological times, we observe Big Bang singularities, solutions where the scale factor asymptotically approaches zero, and Small Bangs. The latter are solutions for which the Hubble parameter diverges at a finite value of the scale factor. They appear generically in our model for certain curvature couplings. An early inflationary era is observed for a specific value of the curvature coupling without further assumptions (unlike in many other inflationary models). A late-time Dark Energy period is present for all solutions with positive cosmological constant, numerically suggesting that a ``cosmic no-hair'' theorem holds under more general assumptions than the original Wald version which relies on classical energy conditions. The classical fields in consideration can be viewed as resembling one-point functions of a semiclassical model, in which the cosmological expansion is driven by a quantum field.

\end{abstract}

\newpage

\tableofcontents

\newpage

\section{Introduction and motivation}
\label{sec:introduction}

Cosmology is the study of largest observable scales, modeling the universe as a whole. Based on observations, the central assumption is that the universe is spatially homogeneous and isotropic. Mathematically, this is modelled by assuming that the spacetime is Friedmann–Lemaître–Robertson–Walker (FLRW), a manifold of the form $M=I_t\times\mathbb R^3$ and thereon has a spacetime metric
\begin{equation}
\label{eq:cosmological_metric}
  g := -\dif t^2 + a(t)^2 g_{\mathbb R^3} \,.
\end{equation}
The function $a$ is called the scale factor and cosmological models are usually formulated in terms of an ODE for $a$. We restrict our work to the case of flat spatial sections. The other two cases of a $3$-sphere or $3$-hyperbolic space are disregarded in this work. Cosmological models are mostly (but not exclusively) obtained from the Einstein equation
\be
\label{eqn:einst}
G_{\mu \nu} +\Lambda g_{\mu \nu}=\kappa T_{\mu \nu}\,,
\ee
where $T_\munu$ is a suitable stress-energy tensor and $G_{\mu \nu}=R_{\mu\nu}-\frac12 R\,g_{\mu\nu}$ is the Einstein tensor. $\Lambda$ is the cosmological constant and $\kappa$ is a parameter. The latter is typically assumed equal to $8\pi G/c^4$, where $G$ is the gravitational constant and $c$ the speed of light in S.I.\ units. 

In this paper, we study classical free real scalar fields, governed by the Klein-Gordon equation
\be
\label{eqn:Klgord_intro}
\left(-\nabla^\sigma \nabla_\sigma+m^2+\xi R \right) \phi=0 
\ee
with mass $m$ and curvature coupling $\xi$.
Particularly, we restrict to fields that are ``cosmological'' in the sense that they are assumed to share the symmetry of the metric in terms of homogeneity and isotropy. Moreover, we allow for a non-zero cosmological constant, which often is absorbed into the RHS of \eqref{eqn:einst} as $T_\munu^\textup{DE}=-\frac{\Lambda}{\kappa} g_\munu$. Viewing $\Lambda$ as a part of $T_\munu$, it is also referred to as Dark Energy.

Klein-Gordon fields can drive the cosmological expansion in the classical or the semiclassical level. In the semiclassical regime, quantum effects are important but the spacetime curvature remains small and thus can be treated classically. The regime is usually described by the semiclassical Einstein equation
\be
\label{eqn:SEE}
G_{\mu \nu}+\Lambda g_{\mu \nu}=\kappa \langle T^{\text{ren}}_{\mu \nu} \rangle_{\omega} \,.
\ee
Here, the RHS of \eqref{eqn:einst} is modeled as the expectation value of the renormalized stress-energy tensor of a quantum Klein-Gordon field in a Hadamard state $\omega$, whereas the LHS is modelled by a classical metric. In that context, homogeneity and isotropy are in close relation with a state being distinguished as a vacuum. 

We refer to the moment approach to the cosmological semiclassicial Einstein equation developed in \cite{Gottschalk:2018kqt}, in which the quantum stress-energy tensor can be split as
\begin{equation}
\label{eq:SCE_with_split_QSE_tensor}
	\big\langle T_\munu^\mathrm{ren}\big\rangle_\omega
	=
	\big\langle T_\munu\big\rangle_{\omega_1}+\big\langle T_\munu\big\rangle_{\omega_2^\mathrm{reg}}+\Theta_\munu^\mathrm{Had}+\Theta_\munu^\mathrm{ta}+\Theta_\munu^\mathrm{ren}(c)\,.
\end{equation}
Here, $\big\langle T_\munu\big\rangle_{\omega_1}$ and $\big\langle T_\munu\big\rangle_{\omega_2^\mathrm{reg}}$ are the contributions of the one-point function and of the regularized two-point function, respectively. $\Theta_\munu^\mathrm{Had}$, $\Theta_\munu^\mathrm{ta}$ and $\Theta_\munu^\mathrm{ren}(c)$ are contributions of the Hadamard condition on the two-point function, of the trace anomaly and of the renormalization freedom, respectively. The latter three are independent of the precise choice of (cosmological Hadamard) state, while the former two resemble this choice.

In \cite{Gottschalk:2021pkr}, Minkowski-like states were constructed fulfilling $\big\langle T_\munu\big\rangle_{\omega_2^\mathrm{reg}}=0$. Under this assumption, \cite{Gottschalk:2021pkr} provides a numerical survey indicating that the three terms $\Theta_\munu^\mathrm{Had}$, $\Theta_\munu^\mathrm{ta}$ and $\Theta_\munu^\mathrm{ren}(c)$ can be effectively modeled by a cosmological constant in the sense of a late-time de Sitter phase. Similar indications have been found earlier in \cite{DEfromQFT,haensel2019}, using other kinds of states for conformally coupled fields $\xi=\frac16$. 

In the same reference it was also assumed that $\big\langle T_\munu\big\rangle_{\omega_1}=0$. However, the contribution of the one-point functions can be modelled by the classical Klein-Gordon field. This is an important motivation to study such fields, develop a systematic approach, and evaluate their effect in the semiclassical context. In future work, we plan to incorporate that effect. 

Additionally, a systematic investigation of backreaction effects of a classical field is interesting in its own right. We note that in order to allow for cosmological solutions of \eqref{eqn:einst}, the stress-energy is necessarily that of a perfect fluid (i.e.\ homogeneous and isotropic). Homogeneity and isotropy appear to be the most natural assumptions on fields that guarantee such a perfect fluid shape (more general fields seem conceivable, however it seems difficult to specify the most general case explicitly). 

Homogeneous and isotropic fields have previously been included in cosmological models, for the most part in the context of inflationary models based on the slow-roll approximation. Inflation refers to the suggested period of exponential expansion of the early universe \cite{Guth:1980zm} and fields causing such a phase are often referred to as inflatons. While inflation has been a successful theory in solving cosmological problems and explaining the emergence of cosmological perturbations \cite{Guth:1982ec, Mukhanov:1990me}, the study of the inflaton is constrained in a regime where certain strong assumptions hold. The slow-roll conditions (see \cite{Riotto:2002yw} for a review) are rather strong, particularly if they should ensure that inflation lasts long enough to solve, for example, the cosmic horizon problem.  

On the other hand, non-minimally coupled fields are rarely taken into account in the literature along with a comprehensive study of the classical Klein-Gordon field in cosmological spacetimes. Here, we explore its effects in full generality, allowing for a non-zero mass, a cosmological constant and general coupling to the curvature. The equations studied, the trace equation derived from \eqref{eqn:einst} and the field equation reveal a rich parameter space with different types of solutions. The system of two differential equations derived is of the form 
\blea
\label{eqn:systsimple}
\ddot{a}&=&f_1(\dot{a}, a, \dot{\phi}, \phi) \\
\ddot{\phi}&=&f_2(\dot{a}, a, \dot{\phi}, \phi) \,,
\elea
where $a(t)$ is the scale factor, $\phi(t)$ the scalar field and $f_{1,2}$ polynomial functions (see \eqref{eqn:syst} for the exact expressions). The system requires four initial conditions $a(0), \dot{a}(0),\phi(0)$ and $\dot{\phi}(0)$. The scale factor at $t=0$ is conventionally set to $1$, while $\dot{a}(0)$ is the Hubble constant today and it fixes the time scale. Reasonable initial values of the field and its derivative are, in principle, unknown. Instead of leaving them as free parameters, we fix them using two constrains: the energy equation (${00}$ component of \eqref{eqn:einst}) and the value of the deceleration parameter today, for which we have observational input. Aside from this observational input (which heavily relies on the structure of the $\Lambda$CDM model), having control over an initial deceleration parameter is generally a desirable property of a cosmological model. Thus, our parameters are the mass of the field $m$, the cosmological constant $\Lambda$ and the coupling constant $\xi$. The initial value of the deceleration parameter $q_0$ is either allowed to vary within a range of reasonable values or it set to its observational $\Lambda$CDM value.

A surprising finding was that not all parameter combinations are allowed. Typically, for a fixed choice of parameters we have no, two or four solutions to the algebraic constraints on the initial values $(\phi(0),\dot{\phi}(0))$ for the field. By the symmetry of the model under replacing $\phi\mapsto-\phi$ we thus have, depending on the parameters, one or at most two inequivalent cosmologies emerging for a fixed choice of parameters.

We are especially interested in the early and late time behavior of these solutions. At early times one of the main motivating questions is if the system originates in a singularity. In the cosmological singularity theorem by Hawking \cite{Hawking:1966sx}, the primordial singularity is defined as timelike geodesic incompleteness without specifying its type or physical properties. Hence, often in cosmology the initial singularity is defined dynamically, as the scale factor becoming zero at a finite time. We indeed observe those kinds of solutions which we call Big-Bang type singularities. But interestingly, we observe another kind of singularity: the scale factor is non-zero but its derivatives diverge at finite time. We call this (seemingly unphysical solution) a Small-Bang type singularity. Both solutions correspond to timelike geodesically incomplete spacetimes but have very different physical consequences. Surprisingly, we discovered that Small-Bang singularities are generic for $0<\xi<1/6$ making them a distinct feature of classical Klein-Gordon cosmologies. Bounce cosmologies, which have been the topic of several works \cite{Battefeld:2014uga} were not observed in our setup. The third type of early times behavior is the one where the scale factor asymptotically approaches zero. For the special value $\xi=\frac3{16}$, the expansion is exactly exponential; an inflationary phase. This phase is observed without approximations and is present even with a zero mass field. We omit to numerically specify the length of this exponential expansion. An approach with more sophisticated numerical algorithms and whether or not such a phase is sufficiently long to solve cosmological problems is left for the future. However, it is notable that this early exponential expansion is also a fairly generic feature for this value of the curvature coupling.

In terms of late times behavior, all the solutions with $\Lambda>0$ studied, show exponential expansion compatible with observations. A ``cosmic no-hair'' theorem \cite{Wald:1983ky}, meaning a de Sitter late time expansion is verified numerically for the cases studied. We note that the classical theorem of Wald \cite{Wald:1983ky} is not applicable due to the non-minimally coupled Klein-Gordon field as it violates the energy conditions this theorem relies on.

Our methods reach from analytically solving ODEs, if possible, over an explicit algebraic approach treating the phase space constraints to, finally, approaching  the dynamical system by numerical methods. For the numerical approach we choose a solver for stiff problems due to our expectation of singularities. Moreover, we develop a point of view on the parameters constraints as conic section equations. Thereby, the dynamical system defined by our model can, to some extend, be understood in terms of a dynamical behavior of the conic sections obeying these constraints. Particularly in asymptotic regimes, we can algebraically identify certain limits of solutions to our model.

The paper is organized as follows: In Section~\ref{sec:flrw} we briefly review the FLRW model and the known solutions with a constant equation of state. We also define the different kinds of cosmological singularities and connect them with the singularity theorems. In Section~\ref{sec:model} we introduce the Klein-Gordon field with generic curvature coupling and setup the cosmological model we study. In Section~\ref{sec:special} we study the Minkowski and de Sitter solutions and how they appear in our model as well as analytic solutions for special values of the parameters. In Sections~\ref{sec:Parameter_constraints} and \ref{sec:evolution} we present a survey on numerical solutions. Section~\ref{sec:Parameter_constraints} focuses on the constraints in parameter space from the energy equation and the initial deceleration parameter. Section~\ref{sec:evolution} includes numerical solutions for the evolution of the scale factor with particular emphasis in the early and late time behavior. We conclude in Section~\ref{sec:discussion} with a summary and discussion of our results as well as future perspectives. 

\medskip

\noindent
\textbf{Conventions:} Unless otherwise specified, we work in $n=4$ spacetime dimensions and use metric signature $(-,+,\dots,+)$. The convention used for the metric, the Riemann tensor and the Einstein Equation is the $(+,+,+)$ according to Misner, Thorne and Wheeler \cite{MTW}.

\section{FLRW cosmology}
\label{sec:flrw}

We begin with a brief review of cosmological spacetimes. Thereafter, the known solutions of the Einstein equation for flat cosmologies with a perfect fluid stress-energy tensor and the $\Lambda$CDM model (the ``standard model of cosmology'') are revisited. We conclude the section with a discussion of the types of cosmological singularities and their connection with singularity theorems. 

\subsection{FLRW spacetimes}

The cosmological principle states that on the largest observable scale our universe is spatially homogeneous and isotropic. Homogeneous and isotropic $3$-manifolds are fully classified by their (constant) curvature, yielding the three cases of flat space $\mathbb R^3$, a $3$-sphere or $3$-hyperbolic space (up to rescaling in the latter two cases). As we mentioned, we study the case of a flat, four-dimensional FLRW spacetime, described by the metric in \eqref{eq:cosmological_metric} with a smooth scale factor $a:I_t\to(0,\infty)$ on some interval $I_t$.

Cosmological models are usually derived from the Einstein equation \eqref{eqn:einst}.  In order to allow for solutions of \eqref{eqn:einst} of the form \eqref{eq:cosmological_metric}, $T_\munu$ is necessarily of the perfect fluid form
\begin{equation}
(T^\mu{}_\nu)=\mathrm{diag}(-\varrho,p,p,p)\,,
\end{equation}
where $\varrho$ and $p$ are called the energy density and pressure, respectively. Moreover, since the same holds true for the LHS of \eqref{eqn:einst}, the stress-energy tensor must be covariantly conserved, $\nabla^\mu T_\munu=0$. We remark that depending on how $T_\munu$ is modelled, this may be automatically fulfilled or it needs to be ensured later. In this work we study classical Klein-Gordon fields that are derived from an underlying action principle and such stress-energy tensors are innately conserved.

In terms of $\varrho$ and $p$, the Einstein equation \eqref{eqn:einst} in the cosmological setting breaks down into a system of two ODEs, namely the energy component ($\mu=\nu=0$) and one of the spatial components (e.g.\ $\mu=\nu=1$). Usually, one passes to the equivalent description of these ODEs by certain linear combinations, such as for instance the Friedmann equations. We will, however, stick to the equivalent description via the energy equation and the trace of the Einstein equation,
\begin{equation}
    \label{eq:energy_and_trace_intro}
    \frac{\dot a^2}{a^2} = \frac\kappa3 \rho+\frac{\Lambda}{3}
    \qquad\textup{and}\qquad
   -6\left(\frac{\ddot a}{a}+\frac{\dot a^2}{a^2}\right)+4\Lambda=\kappa(3p-\varrho)\,,
\end{equation}
respectively, where we used that the Ricci scalar curvature for a metric of the form \eqref{eq:cosmological_metric} specializes into 
\be
R[\,a\,] = 6\left(\frac{\ddot a}{a} + \frac{\dot a^2}{a^2}\right) \,.
\ee
Here, and throughout this article, we use dots to denote derivatives with respect to the cosmological time coordinate $t$. If we, moreover, take into account the conservation condition $\nabla^\mu T_\munu=0$, the energy equation implies the trace equation. Conversely, the trace equation implies the energy equation on any interval where $\dot a\neq 0$, provided that the energy equation holds at one instance of time in that interval (e.g.\ for initial values for a numeric simulation). 
 
Since a particular value of $a$ has no inherent physical content and cosmological models are typically invariant under replacing $a$ by a scalar multiple, it is convenient to study solutions $a$ of a cosmological model in terms of the Hubble parameter
\be
H[\,a\,]:I_t\to\mathbb R,~t\mapsto\frac{\dot a(t)}{a(t)} \,,
\ee
and the deceleration parameter
\begin{equation}
\label{eqn:deccel}
q[\,a\,]:I_t\to\overline{\mathbb R},~t\mapsto-\frac{a(t)\ddot a(t)}{\dot a(t)^2}\,,
\end{equation}
where we set $q[\,a\,](t)=\infty$ if $\dot a(t)=0$ at some $t$. We will use $q$ to define an initial value for $\ddot a$ in an $a$- and $t$-scale invariant way. Also we introduce a shifted version of $q$, namely the equation of state (EOS) coefficient
\begin{equation}
\label{eq:deceleration_state_fraction_correspondence}
\Gamma[\,a\,]=\frac{2}{3}q[\,a\,]-\frac{1}{3}
\end{equation}
and note that whenever $a$ solves \eqref{eq:energy_and_trace_intro} with $\Lambda=0$ and some $\varrho$ and $p$, then
\begin{equation}
\label{eqn:gamma}
\Gamma[\,a\,]=\frac{p}{\varrho}\,.
\end{equation}
The EOS coefficient is often also referred to as barotropic index.
Of particular interest are solutions to \eqref{eq:energy_and_trace_intro} for which $\Gamma[\,a\,]$, or equivalently $q[\,a\,]$, is constant.

\subsection[Constant EOS solutions and \texorpdfstring{$\Lambda$}{Lambda}CDM cosmology]{Constant EOS solutions and \texorpdfstring{$\mathbf{\Lambda}$}{Lambda}CDM cosmology}
\label{sec:constantEOS_and_LCDM_cosmology}
A cosmological model should, given some initial values for the dynamical degrees of freedom at some initial time, return the scale factor as a function of time.  Notice that $\varrho$ is implicitly governed by a differential equation in terms of the conservation condition $\nabla^\mu T_\munu=0$. However, this is not true for $p$ and hence, the system \eqref{eq:energy_and_trace_intro} is not a proper cosmological model yet. We need to model how $p$ depends on $\varrho$.

While one possibility is to model both $\varrho$ and $p$ with one or more degrees of freedom that are governed by their own differential equation (as we do in Section~\ref{sec:The_classical_Klein-Gordon_field}), a simpler yet useful approach is to assume $\varrho:I\to\mathbb R$ as underlying degree of freedom and $p$ to be a scalar multiple of $\varrho$ with constant ratio 
\be
\gamma :=\frac{p}{\varrho} \,.
\ee
Equivalently, we study stress-energy tensors of the form 
\begin{equation}
\label{eq:gamma_type_SE_tensor}
(T[\,\gamma\,]{}^\mu{}_\nu)=\mathrm{diag}(-\varrho,\gamma\varrho,\gamma\varrho,\gamma\varrho)\,,
\end{equation}
which we refer to as $\gamma$-type. The particular value of $\gamma=-1$ is what we call Dark Energy. The values $\gamma=0$ and $\gamma=\frac13$ are often referred to as matter/dust and radiation, respectively. Notice that in the radiation case we have $T[\,\frac13\,]^\mu{}_\mu=0$.

On the level of solutions, assuming for now $\Lambda=0$, when $\gamma$ is constant, from \eqref{eq:deceleration_state_fraction_correspondence} and \eqref{eqn:gamma} we have
\begin{equation}
\label{eq:constant_EOM_effective_model}
\Gamma[\,a\,](t)=\gamma
\qquad\textup{and}\qquad
q[\,a\,](t)=q_0\,.
\end{equation}
From \eqref{eq:deceleration_state_fraction_correspondence} we have
\be
\label{eq:q_in_terms_of_Gamma}
q_0=\frac{3\gamma+1}{2} \,.
\ee
For $\gamma\neq-1$ (equivalently, $q_0\neq-1$) the effective models \eqref{eq:constant_EOM_effective_model} are solved by
\begin{equation}
\label{eq:gammatype_solution}
    a(t)\propto t^{\frac{2}{3(\gamma+1)}}=t^{\frac{1}{q_0+1}}
\end{equation}
as well as time-translations, -reflections and -rescalings thereof.
In the Dark Energy case $\gamma=-1$ (equivalently, $q_0=-1$), in turn, they are solved by
\begin{equation}
    a(t)\propto\exp(Ht)\,.
\end{equation}
Exponential scale factors are referred to as de Sitter expansions of de Sitter solutions (to a certain model). They play an important role in cosmology.

For all $\gamma\in\mathbb R$ we have
\begin{equation}
\label{eq:gamma_type_rho_of_a}
    \varrho(t)\propto\frac{1}{a(t)^{3(\gamma+1)}}\,.
\end{equation}
We note how in the case $\gamma=-1$
\begin{equation}
    \label{eq:cosmological_constant_as_dark_energy}
    T[-1]{}_\munu=-\varrho_0 g_\munu \,,
\end{equation}
with a constant $\varrho_0$. Hence, a Dark Energy-type stress energy tensor can be absorbed into the cosmological constant or vice versa. 

Using \eqref{eq:gamma_type_rho_of_a}, we can now build more sophisticated cosmological models by assuming that multiple types of $\gamma$-matter are driving the cosmological expansion. Formally, for $S\subset\mathbb R$ we set
\begin{equation}
\label{eq:M-type_model}
    T[S]_{00}=\sum_{\gamma\in M}\frac{\Omega_\gamma}{a(t)^{3(\gamma+1)}} \,,
\end{equation}
with some coefficients $\Omega_\gamma>0$ with $\sum_{\gamma\in S}\Omega_\gamma=1$ and insert this into the energy equation \eqref{eq:energy_and_trace_intro}. By \eqref{eq:cosmological_constant_as_dark_energy}, whenever we want to consider a model with $\Lambda>0$, we can instead set $\Omega_{\textup{DE}}:=\Omega_{-1}=\Lambda$ and leave out the cosmological constant. We note that a model with $S=\{-1,\gamma\}$ for one $\gamma\neq-1$ is solved by
\begin{equation}
\label{eq:DE_and_gammatype_solution}
    a(t)\propto\sinh(\beta t)^{\frac{2}{3(\gamma+1)}}
\end{equation}
and time-translations and -reflections thereof, with a suitable $\beta$. In the particular case of $\gamma=\frac13$, that is, $a(t)\propto\sinh(\beta t)^{\nicefrac12}$, we have that $R[\, a \,]$ is constant. Notice that any such solution resembles an approximate late-time de Sitter phase.

An example of such a model is the $\Lambda$CDM model with $S=\{-1,0,\frac13\}$ which satisfies the ODE \eqref{eq:energy_and_trace_intro} is
\begin{equation}
\label{eq:LCDM_ODE}
    \frac{\dot{a}^2}{a^2}~=~H_0^2\left(~\frac{\Omega_\textup{rad}}{a^4}+\frac{\Omega_\textup{dust}}{a^3}+\Omega_\textup{DE}~\right) \,,
\end{equation}
with $a(0)=1$. The $\Lambda$CDM is often deemed the standard model of cosmology and the $\Omega_{\textup{rad}},\Omega_\textup{dust}$ and $\Omega_{\textup{DE}}$ are viewed as coefficients of today's energy distribution in the universe. In \cite{Planck-Collab}, the parameters are determined as 
\begin{equation}
\label{eq:LCDM_parameters}
\Omega_\textup{rad}~=~5.38\cdot10^{-5}~,
    \quad
    \Omega_\textup{dust}~=~0.315~,
    \quad
    \Omega_\textup{DE}~=~0.685
    \quad\textup{and}\quad
    H_0~=~2.19\cdot10^{-18}\,s^{-1}
\end{equation}
by fitting the $\Lambda$CDM model to a observational data from the CMB microwave background. Note that for a solution $a$ of \eqref{eq:LCDM_ODE} with parameters \eqref{eq:LCDM_parameters} one can compute that
\begin{equation}
\label{eq:LCDM_initial_deceleration}
q_{\Lambda\textup{CDM}}:=q[\,a\,](0)=-0.538\,. 
\end{equation}
This value is too sensitive to the precise shape of \eqref{eq:LCDM_ODE} to speak of an experimental value. However, at least it gives us a rough idea of what a realistic initial value for $q_0$ in other cosmological models might be. We also remark that for a solution of \eqref{eq:LCDM_ODE} (or also for the function \eqref{eq:DE_and_gammatype_solution} with $\gamma=\frac13$), $q[\,a\,]$ is a surjective function onto the interval $(-1,1)$.

\begin{figure}
\centering
    \includegraphics[scale=1]{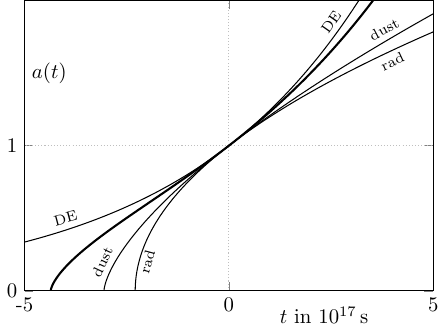}
    \hfill
    \begin{minipage}[b]{.47\textwidth}
\captionsetup{format=plain, labelfont=bf,singlelinecheck=false}
\caption{The graphic shows the evolution of the scale factor $a(t)$ for a radiation-, a dust- and a Dark Energy-driven expansion of an FLRW universe as the thin lines, and the $\Lambda$CDM cosmological model with the parameters \eqref{eq:LCDM_parameters} as the thick line. The Big Bang of the $\Lambda$CDM solution is found at $t_{\textup{BB}}\approx-4.36\cdot10^{17}\,\textup{s}$,  roughly 13.8 billion years.\label{fig:lcdm}
}
\end{minipage}    
\end{figure}

A plot of the evolution of the scale factor for radiation, matter and Dark Energy dominated universes as well as the $\Lambda$CDM model is given in Figure~\ref{fig:lcdm}. We note that all models except the pure cosmological constant one begin with $a \to 0$. At late times, in turn, the Dark Energy dominates and the $\Lambda$CDM model has infinitely accelerated, approximate de Sitter expansion.

\subsection{Types of cosmological singularities}
\label{sec:Types_of_cosmological_singularities}

In this subsection we specify types of singularities for an FLRW spacetime. First, we define explicitly what is commonly referred to as a Big Bang singularity:
\begin{definition}
\label{def:bigbang}
    Let $M=(t^s,T)\times\mathbb R^3$ be a flat FLRW spacetime defined by the scale factor $a:(t^s,T)\to(0,\infty)$ and let $\gamma\in \mathbb R$. $M$ is said to have a $\gamma$-type Big Bang singularity at $t=t^s$ if 
    \begin{equation}
        a(t)\to 0\,,
        \quad
        \dot a(t)\to\infty
        \qquad\textup{and}\qquad
        \Gamma[\,a\,](t)\to\gamma
    \end{equation}
as $t\to t^s$.
\end{definition}
In the literature, usually, the condition of $\varrho(t)\to\infty$ is included into the definition of a Big Bang.
However, by dint of the energy equation \eqref{eq:energy_and_trace_intro}, our definition imposes the latter condition. By \eqref{eq:deceleration_state_fraction_correspondence}, the condition $\Gamma[\,a\,](t)\to\gamma$ can be equivalently formulated in terms of $q[\,a\,](t)\to q^s$, for $q^s$ as given in terms on $\gamma$ by \eqref{eq:q_in_terms_of_Gamma}. We do not observe solutions with $a \to 0$ and  $\Gamma[\,a\,](t)\to \pm \infty$.

Obviously, the cosmological spacetimes defined by the scale factors \eqref{eq:gammatype_solution} and \eqref{eq:DE_and_gammatype_solution} have $\gamma$-type Big Bang singularities if $\gamma>-\frac13$ or $q^s>0$ (where we violate the $\dot a(t)\to\infty$ condition for $\gamma\in(-1,-\frac13]$/$q^s\in(-1,0]$). Also, given some finite $S\subset\mathbb R$ with $\max (S)>-\frac{1}{3}$ one would expect that the $\max (S)$-type matter in the energy equation defined by \eqref{eq:M-type_model} is dominant in the limit $a(t)\to 0$. 

We also define the following type of singularity:
\begin{definition}
\label{def:smallbang}
    Let $M=(t^s,T)\times\mathbb R^3$ be a flat FLRW spacetime defined by the scale factor $a:(t^s,T)\to(a_{\textup{SB}},\infty)$ and let $a_{\textup{SB}}>0$. $M$ is said to have a Small Bang singularity in $t=t^s$ if 
    \begin{equation}
        a(t)\to a_{\textup{SB}}
        \qquad\textup{and}\qquad
        \dot a(t)\to\infty
    \end{equation}
    as $t\to\infty$.
\end{definition}

Again, using the energy equation \eqref{eq:energy_and_trace_intro} a Small Bang singularity imposes $\varrho(t)\to\infty$. A similar type of singularity which is extensively studied in the literature is the so-called quiescent or sudden singularity (e.g.\ \cite{Andersson:2000cv}). However, these singularities are usually only called that if $\varrho$ remains bounded or even, in so-called generalized sudden singularities, also $p$ is bounded, but a higher derivative of them diverges. Notice that also for Small Bang singularities one can assign an asymptotic dominant behavior by studying $\Gamma[\,a-a_{\textup{SB}}\,]$/$q[\,a-a_{\textup{SB}}\,]$. While spacetimes admitting a Small Bang singularities are most probably not physically realized, we will see how they occur generically in classical Klein-Gordon cosmology. 

Finally, we should make a connection with the singularity theorems. First of all, the singularity theorems introduced by Penrose \cite{Penrose:1964wq} and Hawking \cite{Hawking:1966sx} define a singularity as geodesic incompleteness. The theorem applicable in cosmology is the Hawking one which predicts timelike geodesic incompleteness. Both types of singularities defined here lead to geodesic incompleteness and the theorem does not distinguish between them. However, the original Hawking singularity theorem cannot be applied in $\Lambda$CDM cosmology. The reason is one of the assumptions of the theorem, namely the timelike convergence condition. The condition states that the Ricci tensor contracted with two unit timelike vectors $t^\mu$ is everywhere non-negative:
\be
R_{\mu \nu} t^\mu t^\nu \geq 0 \,.
\ee
Using the Einstein equation, this condition is translated into a condition for the matter content of the model \cite{Kontou:2020bta}, namely the strong energy condition, which for $4$ spacetime dimensions takes the form:
\be
\left(T_{\mu \nu}-\frac{1}{2}g_{\mu \nu}T^\sigma{}_\sigma \right)t^\mu t^\nu \geq 0 \,.
\ee
For a perfect fluid stress-energy tensor, the condition requires $\varrho+p \geq 0$ and $\varrho+3p \geq 0$. For $\gamma$-type models and assuming that the energy density is non-negative, the condition is violated for any $\gamma < -1/3$, thus any Dark Energy dominated cosmology. We should note here that the violation of the strong energy condition does not mean singularity avoidance but simple non-applicability of the theorem. 

It was recently shown \cite{Brown:2018hym} that the classical Klein-Gordon field, which is the main interest of this work, obeys a weaker, averaged energy condition (for $\xi<1/4$) which is sufficient to prove timelike geodesic incompleteness and can be applied in cosmology. The results of this theorem are in agreement with our work where singularities of one the two types are predicted in all cases, or the scale factor becomes so small that the theorem is again not applicable.

\section{The model}
\label{sec:model}

In this section, we first briefly introduce the classical Klein-Gordon field and then setup the cosmological model we will analyze. 

\subsection{The classical Klein-Gordon field}
\label{sec:The_classical_Klein-Gordon_field}

The classical action integral for a free, real scalar field $\phi:M\to\mathbb R$ in four-dimensions is
\be
\label{eqn:graction}
S_{\textup{KG}}=\frac{1}{2}\int d^4 x \sqrt{-g} \Big(-(\nabla_\mu \phi) (\nabla^\mu \phi)+\xi R\phi^2+m^2 \phi^2 \Big) \,,
\ee
where $m\ge0$ is the mass of the field and $\xi\in\mathbb R$ a dimensionless coupling constant, called the curvature coupling. The conformal coupling, that is, the $\xi$-value for which the field is conformally invariant, depends on the spacetime's dimension and in our setting is $\xi_{\textup{cc}}=\frac16$. Another widely studied case is minimal coupling, $\xi_{\textup{mc}}=0$.

The field equation, that is, the Euler-Lagrange equation for \eqref{eqn:graction} with respect to $\phi$, is
\be
\label{eqn:Klgord}
\left(-\nabla^\sigma \nabla_\sigma+m^2+\xi R \right) \phi=0 \,.
\ee
The stress-energy tensor for $\phi$, on the other hand, is obtained by varying the action with respect to $g_\munu$ and is
\begin{align}
\label{eqn:tmn}
T_{\mu \nu}&=(1-2\xi)(\nabla_\mu \phi)(\nabla_\nu \phi)- \frac12 (1-4\xi) g_{\mu\nu} (\nabla^{\sigma} \phi) (\nabla_{\!\sigma} \phi) - \frac12 g_{\mu\nu} m^2 \phi^2  \notag\\
&\hspace{6cm}+ \xi \bigr(G_{\mu\nu} \phi^2 - 2 \phi \nabla_{\!\mu} \nabla_{\!\nu} \phi + 2 g_{\mu \nu} \phi \nabla^{\sigma}\nabla_{\!\sigma} \phi \bigr) \,.
\end{align}
We notice that its trace is given by
\begin{equation}
\label{eq:KG_stress_energy_tensor_general}
T^\mu{}_\mu=(6\xi-1)\big((\nabla^\mu\phi)(\nabla_{\!\mu}\phi)+\phi\nabla^\mu\nabla_{\!\mu}\phi\big)-m^2\phi^2\,,
\end{equation}
where $T^\mu{}_\mu=0$ for a massless, conformally coupled field.

Finally, we remark that while we restrict ourselves to the Klein-Gordon field as defined by \eqref{eqn:graction}, the action of a scalar field is often written more generally in terms of a potential $V(\phi)$. Our case emerges by considering $V(\phi)=\frac12m^2\phi^2+\frac12\xi R\phi^2$. Then, if one allows $V$ to have a constant offset $V(0)>0$, this offset enters the stress-energy-tensor and hence the Einstein equation as effective cosmological constant. Vice versa, in analogy with the $M$-type stress-energy tensors \eqref{eq:M-type_model}, a non-vanishing cosmological constant $\Lambda$ can be absorbed into the stress-energy tensor as a constant contribution in $V$.

\subsection{Setup of cosmological solutions}
\label{sub:setupsolutions}

We want to study cosmological solutions to the Einstein equation where gravity is sourced by a Klein-Gordon field. Hereby, we assume that the field is spatially homogeneous and isotropic and hence, simply a function of time. 

Formally, we study pairs of functions $(a,\phi)$ on an interval $I_t$ such that the flat FLRW metric defined by $a$ (cf.\ \eqref{eq:cosmological_metric}) and the field $\phi$ obey the Einstein equation
\begin{equation}
\label{eq:Einstein_equation_general}
G_\munu+\Lambda g_\munu=\kappa T_\munu
\end{equation}
with $T_\munu$ from \eqref{eqn:tmn} coupled to the Klein-Gordon equation \eqref{eqn:Klgord}. As usual in cosmology, the full Einstein equation is equivalent with its energy component and its trace,
\begin{equation}
\label{eq:trace_and_energy_equation}
G_{00}+\Lambda g_{00}=\kappa T_{00}
\qquad\textup{and}\qquad
-R+4\Lambda=\kappa T^\mu{}_\mu\,,
\end{equation}
respectively. Moreover, covariant conservation of $T_\munu$, that is $\nabla^\mu T_\munu=0$, implies that the energy equation alone is sufficient to solve the full Einstein equation (whenever $\dot a\neq 0$). On the other hand, it is often favourable to study the trace equation and view the energy equation as a constraint for the choice of initial values. This is also sufficient for solving the full Einstein equation (whenever $\dot a\neq 0$) and in many models it is numerically more feasible.

In the cosmological setting the energy, trace and Klein-Gordon equation specialize into
\bml
\label{eqn:KG_cosmology}
\begin{eqnarray}
0&=&
	\dot{\phi}^2
		+12\xi\frac{\dot{a}}{a}\phi\dot{\phi}+m^2\phi^2
	+6\xi\frac{\dot{a}^2}{a^2}\phi^2-6\,\frac{\dot{a}^2}{a^2}+2\Lambda\,,
 \label{eq:KG_cosmology_energy_equation}
\\
0&=&
6\left(1+\xi(6\xi-1)\phi^2\right) \left(\frac{\ddot a}{a}+\frac{\dot a^2}{a^2}\right)-4\Lambda
+(1-6\xi)\dot{\phi}^2
+(6\xi -2)m^2\phi^2\,,
\label{eq:KG_cosmology_trace_equation}
\\
	0&=&
	\ddot{\phi}
	+3\frac{\dot{a}}{a}\dot{\phi}+m^2\phi+6\xi\left(\frac{\ddot{a}}{a}+\frac{\dot{a}^2}{a^2}\right)\phi\,,
 \label{eq:KG_cosmology_KG_equation}
\end{eqnarray}
\eml
respectively. Notice that in comparison with \eqref{eq:trace_and_energy_equation} we have set $\kappa=1$, or simply redefine $\phi\mapsto\kappa^{-\nicefrac12}\phi$ under which our system of equations is invariant.

Another useful representation for the trace and the Klein-Gordon equation is
\bml
\label{eqn:syst}
\begin{eqnarray}
\label{eqn:syst1}
\ddot{a} &=&-\frac{\dot{a}^2}{a}+\frac{a}{6}\left(\frac{4\Lambda+  (2-6\xi)m^2\phi^2+(6\xi-1)\dot{\phi}^2}{1+ \xi(6\xi-1)\phi^2}\right) \,,
\\
\label{eqn:syst2}
\ddot{\phi}&=& -3\frac{\dot{a}}{a}\dot{\phi}-\phi \left( \frac{4\Lambda \xi +m^2+ m^2 \xi \phi^2+ \xi(6\xi-1)\dot{\phi}^2}{1+ \xi(6\xi-1)\phi^2}\right) \,,
\end{eqnarray}
\eml
that is, in terms of components of the dynamic vector field on the phase space defined by the independent dynamic variables $(a,\dot a,\phi,\dot\phi)$. Notice that if $\xi(6\xi-1)<0$, or equivalently $\xi_{\textup{mc}}<\xi<\xi_{\textup{cc}}$, we have a critical value $\phi_{\textup{crit}}(\xi)$ which we define as the positive root of
\begin{equation}
\label{eq:phicrit_defining_equation}
1+\xi(6\xi-1)\phi^2_{\textup{crit}}(\xi)=0\,.
\end{equation}
The occurrence of such a singularity is not apparent in the energy equation, although by the above argument the latter equation is already sufficient to determine the dynamics. However, the occurrence of $\ddot a$ in \eqref{eq:KG_cosmology_KG_equation} necessitates taking a derivative of the energy equation which, in the end, introduces the same singularity. Hence, this is not an artifact of choosing to govern the dynamics by the trace equation. 

There is an interesting observation regarding this critical value of $\phi$. First we note that the Klein-Gordon field's stress-energy tensor \eqref{eqn:tmn} also contains a term proportional to the Einstein tensor. Thus, we can rewrite the Einstein equation \eqref{eq:Einstein_equation_general} as
\be
\label{eqn:eeeff}
G_{\mu \nu}=\kappa T^{\text{eff}}_{\mu \nu} \,,
\ee
with an effective stress-energy tensor
\bea
\label{eqn:Teff}
T^{\text{eff}}_{\mu \nu}&=&\frac{1}{1-\xi \phi^2}\bigg( (1-2\xi)(\nabla_\mu \phi)(\nabla_\nu \phi)- \frac12 (1-4\xi) g_{\mu\nu} (\nabla^{\sigma} \phi) (\nabla_{\!\sigma} \phi) - \frac12 g_{\mu\nu}\left( m^2 \phi^2 + \frac{2 \Lambda}{\kappa}\right) \notag\\
&& \hspace{6cm}+ \xi \bigr(- 2 \phi \nabla_{\!\mu} \nabla_{\!\nu} \phi + 2 g_{\mu \nu} \phi \nabla^{\sigma}\nabla_{\!\sigma} \phi \bigr) \bigg) 
\eea
(also absorbing the cosmological constant into $T_\munu^{\textup{eff}}$). The prefactor can be interpreted as altering the Newton constant. $\phi$-values resulting in a non-positive such effective Newton constant have been discussed in the literature and such regimes were discarded as unphysical \cite{Barcelo:2000zf, Fliss:2023rzi}. The claim of a positive effective Newton constant introduces another critical value (for $\xi>0$), equal to $\xi^{-1/2}$, which, for $0<\xi<\frac16$, is inherently smaller than $\phi_{\textup{crit}}(\xi)$ from \eqref{eq:phicrit_defining_equation}. 

One can consider a cosmology starting from values of $\phi$ larger than the value (negative effective Newton constant). There are arguments \cite{Fliss:2023rzi} that show the instability of such models so this cosmological regime also seems unphysical. Thus, starting from an initial value of $\phi$  smaller than both critical values will first approach $\xi^{-1/2}$ and then $|\phi_{\text{crit}}|$, casting doubts to the physical realization of this value. We should note that the minimum value of $|\phi_{\text{crit}}|$ is of Planck scale which is, however, a scale of many inflationary models \cite{Linde:1983gd}.

In turn, we adopt here the viewpoint that the analog of a $\phi$-dependent effective Newton constant is 
\begin{equation}
\label{eq:our_kappa_eff}
    \kappa_{\textup{eff}}=\frac{\kappa}{1+\xi(6\xi-1)\phi^2}\,,
\end{equation}
which more precisely tracks the curvature terms contained in $T_\munu$, besides the obvious one proportional to $G_\munu$. In this expression, also second-derivative terms $g_\munu\,\phi\nabla^\sigma\nabla_{\!\sigma}\phi$ and $\phi\nabla_{\!\mu}\nabla_{\!\nu}\phi$ in the second line of \eqref{eq:KG_stress_energy_tensor_general} (or of \eqref{eqn:Teff}) are taken into account.  These implicitly depend on the background curvature or, at least, on the second derivatives of the scaling factor. Using the Klein-Gordon equation \eqref{eq:KG_cosmology_KG_equation}, the occurring $\ddot\phi$-terms can be replaced and the appearing $R[\,a\,]$-terms can also be collected on the curvature side of the Einstein equation. On the level of the Einstein equation's trace, the resulting prefactor of $R[\,a\,]$ is precisely the denominator in \eqref{eq:our_kappa_eff}. Notice that concerning the term $\phi\nabla_{\!\mu}\nabla_{\!\nu}\phi$, our discussion relies on the assumption of homogeneity and isotropy of $\phi$, however, on the level of the trace $T^\mu{}_\mu$ our argument remains valid regardless of this assumption.

As we will demonstrate in Section~\ref{sec:Phase_space_constraints_from_the_energy_equation}, in our model (with $\Lambda\ge0$) the unphysical event of $\kappa_{\textup{eff}}$ changing its sign cannot happen since a field value $|\phi|>\phi_{\textup{crit}}(\xi)$ contradicts the energy equation \eqref{eq:KG_cosmology_energy_equation} (up to spurious but trivial cases that we also comment on in Section~\ref{sec:Phase_space_constraints_from_the_energy_equation}).

Finally, we compute
\begin{equation}
\label{eq:EOS_coefficient_general}
\frac{p}{\varrho}=\frac{T^j{}_j}{T^0{}_0}=\frac{
		\big(1-4\xi\big) \dot\phi^2 
		- \big(1-4\xi\big) m^2 \phi^2 
		- 4 \xi\frac{\dot a}{a}\phi  \dot\phi
		+2\xi\frac{\dot a^2}{a^2}\phi  ^2 
		+4\xi\big(6\xi-1\big)\big(\frac{\ddot{a}}{a}+\frac{\dot{a}^2}{a^2}\big)\phi^2 
	}{
		\dot{\phi}^2
		+m^2\phi^2
		+12\xi\frac{\dot{a}}{a}\phi\dot{\phi}
		+6\xi\frac{\dot{a}^2}{a^2}\phi^2
	}
\end{equation}
(with a spatial index $j$). Particularly, a minimally coupled field $\xi=0$ with
\begin{equation}
\label{eq:slow-roll_EOS_discussion}
    \Gamma[\,a\,]=\frac{p}{\varrho}=\frac{\dot\phi^2-m^2\phi^2}{\dot\phi^2+m^2\phi^2}
\end{equation}
admits two regimes $\Gamma[\,a\,]\approx 1$ and $\Gamma[\,a\,]\approx-1$ when the kinetic or potential terms are dominant, respectively. Such fields are used in the context of slow-roll inflation models.
Moreover, these different regimes as in \eqref{eq:slow-roll_EOS_discussion} can easily be manipulated by allowing more general potentials as discussed in the end of the previous section. For instance, a potential of the form $V(\phi)=\phi^6-3\phi^4+3\phi^2$ admits a plateau around $\phi=\pm1$ where one would expect $\Gamma[\,a\,]\approx-1$. Homogeneous, isotropic fields in such kind of potentials are used in so-called slow-roll inflation models \cite{Riotto:2002yw}. 

We should mention the inflationary models that use non-minimal coupling but not a self-interacting potential $V(\phi)$. If the potential vanishes identically (still allowing a curvature coupling), as it is the case for the free massless scalar field, then inflation is impossible for $\xi \leq \xi_{\text{cc}}$ \cite{Faraoni:2000wk}. Inflation seems possible in the $V(\phi)=0$ case for strong coupling $|\xi| \gg 1$ (e.g. \cite{Bassett:1997az}). One can transform from non-minimal coupling (Jordan frame) to minimal coupling (Einstein frame) with a conformal transformation and a field redefinition. For $V(\phi) \neq 0$ the frame transformation leads to a different potential for the Einstein frame. Some authors have used this technique to generate inflation from non-minimal coupling \cite{Faraoni:1998qx}. However, care is needed in this process as the equivalence of the two frames is not clear \cite{Faraoni:1998qx, Fliss:2023rzi}. Finally, as early inflation requires large field values, one should take into account the critical values mentioned earlier in this subsection, something not done consistently in the literature.

\section{Special solutions}
\label{sec:special}

For some special values of the parameters, analytical solutions to \eqref{eqn:syst} can be found. First, we discuss the class of Minkowski and de Sitter solutions and then the special cases of massless field, conformal coupling and zero cosmological constant. 

\subsection{Minkowski and de Sitter solutions}
\label{sec:Minkowski_and_De_Sitter_solutions}

An obvious class of solutions are the vacuum solutions with $\phi=\dot\phi=0$, namely, the Minkowski solutions for $\Lambda=0$, $a(t)=a_0>0$ for all $t\in\mathbb R$, as well as the de Sitter solutions for $\Lambda>0$, with 
\be
H_{\textup{vac}}(\Lambda):=\sqrt{\frac\Lambda3} \quad \text{and} \quad a(t)=a_0\mathrm{e}^{H_{\textup{vac}}(\Lambda)t}
\ee
for all $t\in\mathbb R$. For $\Lambda<0$, there is no solution with $\phi(t)=0$, which is known as the fact that our choice of flat spatial sections $\mathbb R^3$ is incompatible with an anti-de Sitter metric.

If, in turn, we assume $a$ to be the Minkowski solution in the first place, $a(t)=1$ for all $t\in\mathbb R$, the system of the energy, the trace and the Klein-Gordon equation \eqref{eqn:KG_cosmology} specializes into
\begin{align}
0&=\dot\phi^2+m^2\phi^2+2\Lambda\,,
\\
0&=
(6\xi-1)\dot{\phi}^2
+(2-6\xi )m^2\phi^2
+4\Lambda\,,
\\
0&=\ddot{\phi}+m^2\phi\,.
\end{align}
Assuming that $m>0$ and noting that any solution of the Klein-Gordon equation in this case is  of the form
\begin{equation}
\phi(t)=\phi_0\cos(m\,t)\,,
\end{equation}
 (or a time translation thereof), $\phi_0\in\mathbb R$, we can see that the system of equations above has no solution for any choice of $\phi_0$, $\xi$ and $\Lambda$.

However, for $m=0$ any solution of the Klein-Gordon equation is of the form
\be
\phi(t)=\phi_1\,t+\phi_0\,,
\ee
$\phi_0,\phi_1\in\mathbb R$. The other two equations imply that 
\be
\Lambda=-\frac{\phi_1^2}{2}
\qquad\textup{and}\qquad
0=3(2\xi-1)\phi_1^2\,.
\ee
Thus, a family of solutions for arbitrary $\xi$ and $m=\Lambda=0$ is given by any constant $\phi(t)=\phi_0$. For a non-constant field, we need $\xi=1/2$ and 
\be
\phi(t)=\pm\sqrt{-2\Lambda}\,t+\phi_0 \,,
\ee
with arbitrary $\phi_0$. Physically speaking, a negative cosmological constant is usually associated with a contracting or recollapsing universe. However, here, the massless Klein-Gordon field with $\xi=\frac12$ is capable of cancelling this effect, leading to a constant $a$. 

Next, we assume the underlying spacetime to be a pure de Sitter expansion, $a(t)=\exp(H\,t)$ for all $t\in\mathbb R$, with $H>0$. De Sitter contractions ($H<0$) are regarded using the time reversal symmetry. The system of equations specializes into 
\bml
\label{eq:de_sitter_system}
\begin{eqnarray}
    0&=&
	\dot{\phi}^2
		+12\xi H\phi\dot{\phi}+(m^2
	+6\xi H^2)\phi^2-6H^2+2\Lambda\,,
 \label{eq:de_sitter_energy}
 \\
0&=&(6\xi-1)\dot\phi^2+\Big[(1-6\xi)12\xi H^2+(2-6\xi)m^2\Big]\phi^2-12 H^2+4\Lambda\,,
 \label{eq:de_sitter_trace}
 \\
 0&=&\ddot{\phi}+3H\dot{\phi}+(m^2+12\xi H^2)\phi\,,
 \label{eq:de_sitter_KG}
\end{eqnarray}
\eml
a system of ODEs for the single dynamic degree of freedom $\phi$. The Klein-Gordon field is governed by a damped harmonic oscillator equation \eqref{eq:de_sitter_KG} with an effective de Sitter mass 
\be
M^2=m^2+12\xi H^2\,.
\ee
We note that $M^2$ may also take negative values, however, we keep denoting it as a square in accordance with various other occurrences throughout the literature.

The roots of the characteristic polynomial of $\eqref{eq:de_sitter_KG}$ are
\begin{equation}
\label{eq:charpol_roots_KG_on_dS}
\lambda_\pm=-\frac{3H}{2}\pm\frac{1}{2}\sqrt{9H^2-4M^2}
\end{equation}
and the general solutions in the underdamped ($9H^2<4M^2$), critically damped ($9H^2=4M^2$) and overdamped ($9H^2>4M^2$) regimes are given by
\blea
    \phi_{\mathrm{ud}}(t)&=&c\;\mathrm{e}^{-\frac32Ht}\cos\Big(\tfrac12\sqrt{4M^2-9H^2}\,t\Big)\,,
    \label{eq:underdamped_harmosc_solution}
    \\
    \phi_{\mathrm{cd}}(t)&=&\mathrm{e}^{-\frac32Ht}(c_1+c_2t)\,,
    \\
    \phi_{\mathrm{od}}(t)&=&\mathrm{e}^{-\frac32Ht}\Big(c_1\,\mathrm{e}^{\tfrac12\sqrt{9H^2-4M^2}\,t}+c_2\,\mathrm{e}^{-\tfrac12\sqrt{9H^2-4M^2}\,t}\Big)\,,
\elea
respectively, with some constants $c,c_1,c_2$ (and a time-offset in $\phi_{\mathrm{ud}}$ that we absorb via the time translation invariance of our system). The solutions, $c=0$ or $c_1=c_2=0$, with $H=H_{\textup{vac}}(\Lambda)$ are the vacuum solutions.

In the underdamped case, there are no non-trivial solutions (triviality refers to $c=0$ or $H=0$). However, in a later section we will numerically find solutions that are approximately of the shape $\phi_{\textup{ud}}$ on an asymptotic late-time de Sitter phase.

In the critically damped case, there are no solutions with $c_2 \neq 0$. The assumption $c_2=0$, in turn, implies $m=0$ (if we claim $c_1\neq0$). A vanishing mass and the condition of the critically damped case imply $\xi=\frac3{16}$, a value slightly larger than conformal coupling. Then $H=H_\textup{vac}(\Lambda)$ and the energy and trace equation are fulfilled. In this setting we, indeed have a de Sitter solution with a field of the form $\phi(t)=\textup{e}^{-\frac32H_\textup{vac}(\Lambda)t}$.

In the overdamped case, considering first the massless subcase, we find solutions only for minimal or conformal coupling, for both values resembled by vacuum solutions $H=H_\textup{vac}(\Lambda)$. In the massive overdamped case, we find a variety of solutions, however, all of them are vacuum solutions with $H = H_{\text{vac}}(\Lambda)$. In fact, one can show in a straight-forward computation that non-vacuum solutions with $H \neq H_{\text{vac}}(\Lambda)$ cannot exist.

To conclude, there are no non-trivial de Sitter solutions except the vacuum solution.

Finally, we remark that by a reparameterization $t=\frac1H\log(a)$ we reproduce the bounds 
\blea
    \phi_{\mathrm{od}}(a)&\in &\mathcal{O}\Big(a^{-\frac32+\frac12\sqrt{9H^2-4M^2}}\,\Big)\,,
    \\
    \phi_{\mathrm{cd}}(a)&\in &\mathcal{O}\big(a^{-\frac32}\log(a)\big)\,,
    \\
    \phi_{\mathrm{ud}}(a)&\in &\mathcal{O}\big(a^{-\frac32}\big)
\elea
of solutions solely of the Klein-Gordon equation found in \cite{Natario:2019sap} that were found under very different assumptions on the field.

\subsection{Special values of the parameters}
\label{sec:Special_values_of_the_parameters}

First, we should briefly comment on the simple and probably most discussed model with a minimally coupled, massless field $m=\xi=0$. From \eqref{eq:slow-roll_EOS_discussion}, we read off that such fields impose $\frac{p}{\varrho}=1$. Hence, it can be effectively be described by a stress-energy tensor of the form $T[\,1\,]_\munu$ or $T[\{-1,1\}]_\munu$ (cf.\ \eqref{eq:gamma_type_SE_tensor} and \eqref{eq:M-type_model}), depending on whether $\Lambda=0$ or $\Lambda>0$, respectively. The cosmological expansion driven by such a field is then given by
\begin{equation}
    a(t)\propto t^{\nicefrac13}
    \qquad\textup{or}\qquad
    a(t)\propto \sinh(\beta t)^{\nicefrac13}\,.
\end{equation}

Also, we comment on the $m=\Lambda=0$ and $\xi=1/6$ model. The Klein-Gordon and the trace equation simplify into
\be
\ddot{\phi}= -3\frac{\dot{a}}{a}\dot{\phi}  \,, \qquad \ddot{a} =-\frac{\dot{a}^2}{a}\,,
\ee
respectively, and can be solved analytically. As the model is $a$- and $t$-scale invariant we can, without loss of generality, impose initial values $a(0)=\dot a(0)=1$ (disregarding the Minkowski solution) and obtain 
\be
\label{eq:special_case_confcoup_massless_Lambda0_scale_factor}
a(t)=\sqrt{1+2 t} \,.
\ee
We remark that \eqref{eq:EOS_coefficient_general} specializes into
\begin{equation}
    \frac{p}{\varrho}=\frac13\,\frac{(\dot\phi-H\phi)^2}{(\dot\phi+H\phi)^2}\,.
\end{equation}
Consequently, if either of the terms $\dot\phi$ or $H\phi$ dominates we expect $\Gamma[\,a\,]\approx\frac13$ and hence \eqref{eq:special_case_confcoup_massless_Lambda0_scale_factor} to hold.
The initial values for the field must satisfy the energy equation
\be
0=
	\dot{\phi}^2
		+2\frac{\dot{a}}{a}\phi\dot{\phi}
	+\frac{\dot{a}^2}{a^2}\phi^2-6\,\frac{\dot{a}^2}{a^2}\,.
\ee
Solving the quadratic equation gives
\be
\phi'(0)=\pm \sqrt{6} - \phi(0) \,.
\ee
Therefore, the field evolution is given by
\be
\label{eq:special_case_confcoup_massless_Lambda0}
\phi(t)=\frac{(\pm \sqrt{6}+\phi(0))}{\sqrt{1+2 t}}\mp \sqrt{6} \,.
\ee
Hereby, the initial value $\phi(0)$ is a free parameter, the only one for this simple model. 

Next, we set $m=\Lambda=0$ but leave $\xi$ undetermined. In this case, our model specializes into
\bml
\label{eqn:massless_zerolambda}
\begin{eqnarray}
\label{eqn:massless_zerolambda_trace_eqaution}
\ddot{a}=&-\frac{\dot{a}^2}{a}+\frac{a}{6}\,
\frac{
(6\xi-1)\dot{\phi}^2
}{1+(6\xi-1)\xi\phi^2}
\\
\label{eqn:massless_zerolambda_KG_equation}
\ddot{\phi}=&-3\frac{\dot{a}}{a}\dot{\phi}-
\phi\,\frac{\xi(6\xi-1)\dot{\phi}^2
}{1+(6\xi-1)\xi\phi^2}
\end{eqnarray}
\eml
with the constraint
\begin{equation}\label{eqn:energy_massless_model}
	0=
	\dot{\phi}^2
		+12\xi\frac{\dot{a}}{a}\phi\dot{\phi}
	+6\xi\frac{\dot{a}^2}{a^2}\phi^2-6\,\frac{\dot{a}^2}{a^2}\,.
\end{equation}
Reading \eqref{eqn:energy_massless_model} as a second-order polynomial for $\dot\phi$, we have to claim that the polynomial discriminant is non-negative in order for $\dot\phi$ to be real-valued. This provides us a bound on $\phi$-values as
\begin{equation}\label{eqn:bound_on_phi_values}
    \xi\big(6\xi-1\big)\phi^2
		+1
		\ge0\,,
\end{equation}
that is, the denominator in the dynamic vector field's components in  \eqref{eqn:massless_zerolambda} is non-negative. Applying the same argument and viewing \eqref{eqn:energy_massless_model} now as a second-order polynomial for $\phi$, we obtain the bound 
\begin{align}
    \frac{a^2}{\dot{a}^2}\Big(\xi-\frac16\Big)\dot{\phi}^2
	+1
	&\ge0\,.
\end{align}
Rewriting the trace equation for the initial values for $\phi$ and $\dot\phi$ in terms of $q_0$, we obtain 
\begin{align}
1-q_0&=\frac{1}{6}\frac{a(0)^2}{\dot{a}(0)^2}~
\frac{
(6\xi-1)\dot{\phi}(0)^2
}{1+(6\xi-1)\xi\phi(0)^2}
\end{align}
for our currently discussed parameter choices. In particular, in order to obtain real-valued solutions for $\phi(0)$ and $\dot\phi(0)$ we have to choose $q_0<1$ if $\xi>\frac16$, $q_0=1$ if $\xi=\frac16$ and $q_0>1$ if $\xi<\frac16$. 

A particular class of solutions in our presently discussed parameters setting is given by pure radiation expansion with constant field,
\begin{equation}
\label{eq:radiation_solutions_when_m_equals_lambda_equals_0}
\phi(t)=\phi_0
    \qquad\textup{and}\qquad
      a(t)=\sqrt{1+2t} \,,
\end{equation}
that solve the trace and the Klein-Gordon equation, \eqref{eqn:massless_zerolambda}. These kind of solutions exist whenever $\xi>0$ and the energy equation \eqref{eqn:energy_massless_model} constrains the field's value to 
$\phi_0=\pm\frac{1}{\sqrt{\xi}}$. These solutions coincide in the special case $\xi=\frac16$ with the solutions \eqref{eq:special_case_confcoup_massless_Lambda0_scale_factor} choosing $\phi(0)=\phi_0$.

Finally, we discuss the massless conformally coupled case with an undetermined (positive) cosmological constant. Equations~\eqref{eqn:syst1} and \eqref{eqn:syst2} simplify into 
\blea
\label{eq:trace_equation_for_conformally_coupled_case}
\ddot{a} &=&-\frac{\dot{a}^2}{a}+\frac{2}{3} a \Lambda  \,,\\
\ddot{\phi}&=& -3\frac{\dot{a}}{a}\dot{\phi}-\frac{2}{3}\phi \Lambda\,.
\elea
The energy constraint \eqref{eq:KG_cosmology_energy_equation} becomes
\be
3\frac{\dot{a}^2}{a^2}-\Lambda=\frac{1}{2}\left(\dot{\phi}+\frac{\dot{a}}{a}\phi \right)^2 \,.
\ee
From this equation it is obvious that in this case necessarily
\be
\label{eqn:alambdaconst}
3\frac{\dot{a}^2}{a^2} \geq \Lambda\,,
\ee
that is, $H[\,a\,]$ is bounded below by the vacuum value $H_{\textup{vac}}(\Lambda)$. Rewriting the trace equation \eqref{eq:trace_equation_for_conformally_coupled_case} for the initial values in terms of $q_0$, we obtain
\be
\label{eqn:lambdadef}
(1-q_0)\frac{\dot{a}(0)^2}{a(0)^2}=\frac{2\Lambda}{3} \,,
\ee
showing that in this case $q_0\leq 1$. Along with \eqref{eqn:alambdaconst} we also have $q_0\geq -1$. $q_0=1$ implies $\Lambda=0$ and yields the solutions \eqref{eq:special_case_confcoup_massless_Lambda0_scale_factor}. On the other hand, if $q_0<1$, we have the following constraint
\be
\left(\dot{\phi}(0)\pm \sqrt{\frac{2\Lambda}{3(1-q_0)}} \phi(0) \right)^2=2\Lambda\frac{1+q_0}{1-q_0} 
\ee
 on the initial values of $\phi$ and $\dot{\phi}$.
We note that we have a freedom in choosing initial value for $\phi(0)$ and $\dot{\phi}(0)$ for a given set of parameters. 

We recall that the $\sinh^{\nicefrac12}$-type solutions discussed in Section~\ref{sec:constantEOS_and_LCDM_cosmology} (setting $\gamma=\frac13$) solve the trace equation \eqref{eq:trace_equation_for_conformally_coupled_case}. Indeed, both the stress-energy tensor of a massless, conformally coupled Klein-Gordon field and the radiation stress-energy tensor $T[\,\frac13\,]_\munu$ have a vanishing trace. However, the energy equation imposes different constraints on the initial values for $\phi$ and $\dot\phi$ or on $\varrho$, respectively.

\section{Parameter constraints}
\label{sec:Parameter_constraints}

Usually, a system of two second-order ODEs such as the trace and the Klein-Gordon equation requires four initial values. By convention, the initial value for $a$ is set to $1$, while the initial value for $\dot{a}$ only changes the time scale. What remains are the two initial values of the  scalar field $\phi$. However, as we argued in Section~\ref{sub:setupsolutions} the energy equation introduces a constraint that restricts one of these initial values. First, we examine the solutions of the energy equation in phase space to study the allowed trajectories. Next, we study the solutions from a known deceleration parameter which fixes the fourth initial condition of the system. We conclude the section with a reparametrization of the dynamical system in phase space.

\subsection{Phase space constraints from the energy equation}
\label{sec:Phase_space_constraints_from_the_energy_equation}

Throughout the subsection we use $H=\frac{\dot a}{a}$  and recall the symmetry of our model under $H\mapsto-H$. Moreover, we restrict our considerations to $\Lambda\ge0$ and leave out any time arguments of the dynamical variables.

We consider solutions to the energy equation
\begin{equation}
\label{eq:energy_equation_in_phase_space_section}
    0=f(\phi,\dot\phi,H) \,,
\end{equation}
as the zero set of of the function
\be
f:\mathbb R^3\to\mathbb R\,,~(\phi,\dot\phi,H)\mapsto
\dot{\phi}^2
		+12\xi H\phi\dot{\phi}+m^2\phi^2
	+6\xi H^2\phi^2+2\Lambda-6H^2\,.
\ee

First, recall that for $\xi\in(0,\frac16)$ we defined the critical value $\phi_{\textup{crit}}(\xi)$ in \eqref{eq:phicrit_defining_equation} as the (positive) solution of $1+\xi(6\xi-1)\phi^2_{\textup{crit}}(\xi)=0$, that is, where the phase space vector field's component for $\ddot a$ and $\ddot\phi$ are singular, cf.\, \eqref{eqn:syst}. However, if for such $\xi$-values we assume that $|\phi|\ge\phi_{\textup{crit}}(\xi)$, we can conclude $1+\xi(6\xi-1)\phi^2\le0$ and thus
\begin{equation}
\label{eq:discriminant_of_en_eq_for_phidot_estimate}
    6H^2\big(1+\xi(6\xi-1)\phi^2\big)-m^2\phi^2-2\Lambda\le-m^2\phi^2-2\Lambda
\end{equation}
with equality if and only if $|\phi|=\phi_{\textup{crit}}(\xi)$ or $H=0$.
The left hand side of this estimate, in turn, equals the discriminant of $\dot\phi\mapsto f(\phi,\dot\phi,H)$, read as a second order polynomial in $\dot\phi$. For $\Lambda\geq 0$, the discriminant would be negative and thus unphysical if the inequality holds. Consequently, the zero set of $f$ is confined in 
\be
\big(-\phi_{\textup{crit}}(\xi),\phi_{\textup{crit}}(\xi)\big)_\phi\times\mathbb R^2_{(\dot\phi,H)}\subset\mathbb R^3_{(\phi,\dot\phi,H)}
\ee
for $\Lambda>0$ as $\dot\phi\mapsto f(\phi,\dot\phi,H)$ cannot have a real zero outside of this set. For $\Lambda=0$ and, moreover, $m>0$ we also conclude from \eqref{eq:discriminant_of_en_eq_for_phidot_estimate} that there is no solution for $|\phi|\ge\phi_{\textup{crit}}(\xi)$, at most for $H=0$ (which will be studied below). Finally, only for $\Lambda=m=0$, there can be solutions at $|\phi|=\phi_{\textup{crit}}(\xi)$ (and at $|\phi|>\phi_{\textup{crit}}(\xi)$ only at $H=0$, cf.\ below). To see this, we rewrite 
\be
f \big(\pm\phi_{\textup{crit}}(\xi),\dot\phi,H\big)=\big(\dot\phi\pm6\xi H\phi_{\textup{crit}}(\xi)\big)^2
\ee
and find that $f$ vanishes along the lines parameterized by
\begin{equation}
\label{eq:parametrizations_of_lines_at_phi_crit}
\mathbb R\to\mathbb R^3,~H\mapsto\big(\pm\phi_{\textup{crit}}(\xi),\mp\, 6\xi H\phi_{\textup{crit}}(\xi),H\big)\,.
\end{equation}
To conclude, for $0<\xi<\frac16$ the zero set of $f$ is always confined in 
\begin{equation}
\label{eq:confinement_set_for_solution_set}
\Big(\big[-\phi_{\textup{crit}}(\xi),\phi_{\textup{crit}}(\xi)\big]_\phi\times\mathbb R^2_{(\dot\phi,H)}
\Big)~\cup~\Big(
\mathbb R_\phi\times\big\{(0,0)\big\}_{(\dot\phi,H)}\Big)
~\subset~\mathbb R^3_{(\phi,\dot\phi,H)}
\end{equation}
and can only in the special case $\Lambda=m=0$ have points in the first factor's boundary or in the second factor.

\subsubsection{Surface in phase space}

Next, we study the system of four equations given by
\be
f(\phi,\dot\phi,H)=0
\qquad\textup{and}\qquad
\nabla_{\!(\phi,\dot\phi,H)}f(\phi,\dot\phi,H)=0\,.
\ee
We want to examine the surface in $(\phi,\dot\phi,H)$-space defined by \eqref{eq:energy_equation_in_phase_space_section}. The surface is non-smooth at most at the points where $\nabla_{\!(\phi,\dot\phi,H)}f(\phi,\dot\phi,H) = 0$.

Solving the $\partial_{\dot\phi}$-equation for $\dot\phi$ and plugging the result in the in the $\partial_H$-equation we find
\begin{equation}
\label{eq:ness_crit_on_nonsmoothness_of_surface}
0=-12H\left(1+\xi(6\xi-1) \phi^2\right)\,,
\end{equation}
which shows that critical points are at $|\phi|=\phi_{\textup{crit}}(\xi)$ (only in the case $0<\xi<\frac16$) or at $H=0$. 
As we found above, and noting that
\be
    f(\phi,\dot\phi,0)=\dot\phi^2+m^2\phi^2+2\Lambda\,,
\ee
no point with $H=0$ solves the energy equation \eqref{eq:energy_equation_in_phase_space_section} for $\Lambda>0$ and thus the implicit function theorem implies smoothness of the surface. 

If, on the other hand, $\Lambda=0$ but $m>0$ we found above that only $(\phi,\dot\phi)=(0,0)$ solves both \eqref{eq:energy_equation_in_phase_space_section} and \eqref{eq:ness_crit_on_nonsmoothness_of_surface} at $H=0$. Hence, also in this case the implicit function theorem gives smoothness away from $(0,0,0)$. We will argue below (Section~\ref{sec:singular_points}) that in that point the surface is in fact singular.

If $\Lambda=m=0$ and, moreover, $\xi\in\mathbb R\backslash(0,\frac16)$ the only points that solve \eqref{eq:energy_equation_in_phase_space_section} and \eqref{eq:ness_crit_on_nonsmoothness_of_surface} are points of the form $(\phi,0,0)$, $\phi\in\mathbb R$. Also in this case, we comment on below that the surface is singular along these points.

In the last case, $\Lambda=m=0$ and $0<\xi<\frac16$ we found above that any point that solves both \eqref{eq:energy_equation_in_phase_space_section} and \eqref{eq:ness_crit_on_nonsmoothness_of_surface} is either of the form $(\phi,0,0)$, $\phi\in\mathbb R$, or in the ranges of \eqref{eq:parametrizations_of_lines_at_phi_crit}. However, we can compute for the presently discussed $\xi$-values that
\be
\nabla_{\!(\phi,\dot\phi,H)}f\big(\pm\phi_{\textup{crit}}(\xi),\mp\, 6\xi H\phi_{\textup{crit}}(\xi),H\big)=\left(\pm\frac{12H^2}{\phi_{\textup{crit}}(\xi)}\,0,0\right)\,,
\ee
showing that the surface is still smooth along the ranges of  \eqref{eq:parametrizations_of_lines_at_phi_crit} away from $H=0$. Also in this case we will see below that the surface is indeed singular along points of the form $(\phi,0,0)$, $\phi\in\mathbb R$.

To conclude, the surface defined as the zero set of $f$ is singular at most if $\Lambda=0$ and in that case only at $H=0$.

Next, we we rewrite
\be
f(\phi,\dot\phi,H)=\Phi^tA\Phi
\ee
with
\begin{equation}
\label{eq:conic_section_representation_of_energy_equation}
	A:=
	\begin{pmatrix}
    A_1&\begin{matrix}0\\0\end{matrix}\\
	\begin{matrix}0&0\end{matrix}&A_0
	\end{pmatrix}
	:=
	\begin{pmatrix}
	1&6\xi H&0\\
	6\xi H&m^2+6\xi H^2&0\\
	0&0&2\Lambda-6H^2
	\end{pmatrix}
	\qquad\textup{and}\qquad	
	\Phi:=\begin{pmatrix}
	\dot\phi\\\phi\\1
	\end{pmatrix}\,.
\end{equation}
We see that at each fixed $H$ the zero set of $f$ defines a conic section in the $(\phi,\dot\phi)$-plane. The determinant of $A$ reads as 
\be
\det(A)=(2\Lambda-6H^2)(m^2+6H^2 \xi(1-6\xi)) \,,
\ee
and the determinant and trace of $A_1$ read as
\begin{equation}
\label{eq:det_and_trace_for_conic_section_matrix}
\det(A_1)=m^2+6\xi(1-6\xi)H^2
\qquad\textup{and}\qquad
\tr(A_1)=1+m^2+6\xi H^2\,.
\end{equation}
Then, we can determine the shape of the resulting conic section. The cases are presented in Table~\ref{tab:conic}. In the values of the parameters we assume that $H$ increases in the past. For instance if $\det{A} \neq 0$ and $\xi(1-6\xi)<0$, we will always have $\det({A_1})<0$ (hyperbola) for some cosmological time $t<0$.

\begin{table}
    \centering
    \begin{tabular}{|c|c|c|c|} \hline 
         & $\det{A}$ & $\det{A_1}$  & Parameters \\ \hline 
        Hyperbola & $\neq 0$ & $<0$ & $\xi(1-6\xi)<0$ \\ \hline 
        Parabola & $\neq 0$ & $=0$  & Impossible \\ \hline 
        Ellipse & $\neq 0$ & $>0$ & $0\leq \xi \leq 1/6$ (equality for $m\neq 0$) \\ \hline 
       Two intersecting lines  & $=0$ & $<0$ & $H=H_{\text{vac}}$ and  $\xi(1-6\xi)<0$ \\ \hline 
        Two parallel lines & $=0$ & $=0$ & $\xi(1-6\xi)<0$ and $m^2+6H^2 \xi(1-6\xi)=0$ \\ \hline 
       A single point  & $=0$ & $>0$ & $H=H_{\text{vac}}$ and $0\leq \xi \leq 1/6$ (equality for $m\neq 0$) \\ \hline
    \end{tabular}
    \caption{The shape of the conic section for the different values of the determinants of $A$ and $A_1$. In the values of the parameters we take into account that $H$ will become sufficiently large at some cosmological time $t<0$. We note that we cannot have a parabola, as when the determinant of $A_1$ equals zero, so does the determinant of $A$ in which case we have two parallel lines.}
    \label{tab:conic}
\end{table}

\subsubsection{Localization of solutions}

We want to use the perspective on the energy equation \eqref{eq:energy_equation_in_phase_space_section} as a conic section equation to show that if $0<\xi<\frac16$ the solution set of \eqref{eq:energy_equation_in_phase_space_section} comes arbitrarily close to the planes in $(\phi,\dot\phi,H)$ space defined by $|\phi|=\phi_{\textup{crit}}(\xi)$. Therefore, we perform a blow-up substitution $\dot\phi\mapsto H^2\dot\phi$, that is, we study the equation 
\begin{equation}
\label{eq:blowup_substituted_energy_conic_sections}
0=f(\phi,H^2\dot\phi,H)=:\Phi^t\begin{pmatrix}
    \widetilde A_1&\begin{matrix}0\\0\end{matrix}\\
	\begin{matrix}0&0\end{matrix}&A_0
	\end{pmatrix}\Phi\,.
\end{equation}
We diagonalize
\be
\widetilde A_1=\begin{pmatrix}
    H^4&6\xi H^3\\6\xi H^3&m^2+6\xi H^2
\end{pmatrix}
\ee
and find eigenvalues and eigenvectors as
\be
\lambda_\pm
=
\frac{H^4}{2} + 3\xi H^2 + \frac{m^2}{2} \pm \frac{R}2
\qquad\textup{and}\qquad
v_\pm
=
\begin{pmatrix}\widetilde v_\pm\\1\end{pmatrix}
=
\begin{pmatrix}
\frac{H}{12\xi} - \frac{1}{2H} - \frac{m^2}{12\xi H^3} \pm \frac{R}{12\xi H^3}
	\\
	1
\end{pmatrix}\,,
\ee
where
\be
R=\sqrt{H^8 + 12\xi(12\xi - 1) H^6 + 2(18\xi^2-m^2)H^4 + 12\xi m^2H^2 + m^4}\,.
\ee
An asymptotic expansion of $R$ yields
\begin{align}
    \lambda_+&\in H^4+\mathcal{O}(H^2)\,,
    &
    \widetilde v_+&\in\frac{H}{6\xi}+\mathcal{O}(H^{-1})\,,
    \\
    \lambda_-&\in 6\xi(1-6\xi)H^2+\mathcal{O}(H^{-1})\,,\hspace{-1cm}
    &
    \widetilde v_+&\in-\frac{6\xi}{H}+\mathcal{O}(H^{-3})
\end{align}
and thus for the normalized eigenvectors
\be
\frac{1}{\sqrt{\widetilde v_+^2+1}}v_+
\xrightarrow{H\to\infty}
\begin{pmatrix}0\\1\end{pmatrix}
\qquad\textup{and}\qquad
\frac{1}{\sqrt{\widetilde v_-^2+1}}v_-
\xrightarrow{H\to\infty}
\begin{pmatrix}1\\0\end{pmatrix}
\ee
in the limit $H\to\infty$. Rewriting \eqref{eq:blowup_substituted_energy_conic_sections} (for sufficiently large $H$ such that $6H^2-2\Lambda> 0$) into
\begin{align}
1=\frac{\lambda_+}{6H^2-2\Lambda}
\left(
	\frac{\widetilde v_+}{\sqrt{\widetilde v_+^2+1}}\dot\phi+\frac{1}{\sqrt{\widetilde v_+^2+1}}\phi
\right)^2+
\frac{\lambda_-}{6H^2-2\Lambda}
\left(
	\frac{\widetilde v_-}{\sqrt{\widetilde v_-^2+1}}\dot\phi+\frac{1}{\sqrt{\widetilde v_-^2+1}}\phi
\right)^2
\end{align}
we can read off the lengths of the ellipse's semi-major and -minor axes. We notice that for  $0<\xi<\frac16$ indeed both $\lambda_+$ and $\lambda_-$ are positive at sufficiently large $H$ implied by $\det(\widetilde A_1)>0$ and $\tr(\widetilde A_1)>0$. These semi-axes' lengths fulfill 
\be
\sqrt{\frac{6H^2-2\Lambda}{\lambda_+}}\to0
\qquad\textup{and}\qquad
\sqrt{\frac{6H^2-2\Lambda}{\lambda_-}}\to\frac{1}{\sqrt{\xi(1-6\xi)}\,}=\phi_{\textup{crit}}(\xi)
\ee
as $H\to \infty$. In particular, the ellipses solving \eqref{eq:blowup_substituted_energy_conic_sections} come arbitrarily close to the $|\phi|=\phi_{\textup{crit}}(\xi)$-planes as desired. This does obviously not change under reversing the blow-up, as this merely stretches a conic section at some $H$ in the $\dot\phi$-direction.

Another useful blow-up substitution is given by $\dot\phi\mapsto H\phi$. The conic sections solving 
\begin{equation}
    \label{eq:other_blowup_substituted_energy_conic_sections}
0=\frac{1}{H^2}f(\phi,H\dot\phi,H)=\dot{\phi}^2
		+12\xi\phi\dot{\phi}+\Big(6\xi+\frac{m^2}{H^2}\Big)\phi^2
	-6+\frac{2\Lambda}{H^2}
\end{equation}
(with $H\neq0$ and arbitrary $\xi$) become stable as $H\to\infty$, that is, the analogs of $\lambda_\pm$ and $\widetilde v_\pm$ converge. Obviously, the limits are simply obtained by setting $m=\Lambda=0$. In particular, if we study the model with $\Lambda=m=0$, the conic sections solving \eqref{eq:other_blowup_substituted_energy_conic_sections} do not depend on $H$ at all. Notice that this blow-up substitution is particularly useful for studying the limiting behavior of solutions, especially when $H$ becomes large.

\subsubsection{Plots}

\begin{figure}
\centering
\begin{subfigure}[b]{.49\textwidth}
    \includegraphics{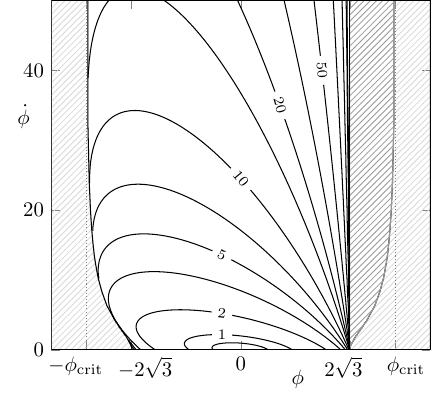}
    \caption{Level sets for $H(\phi,\dot\phi)$}
\end{subfigure}
\hfill
\begin{subfigure}[b]{.49\textwidth}
    \includegraphics{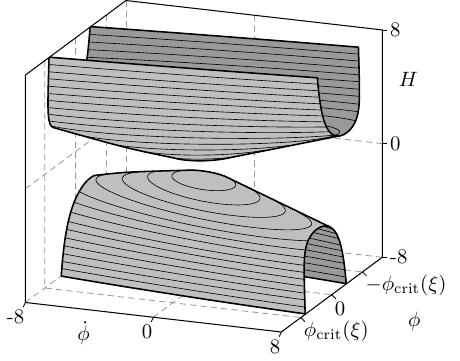}

    \vspace{.3cm}
    \caption{Surface in $(\phi,\dot\phi,H)$-space\qquad}
\end{subfigure}

\smallskip
\begin{minipage}{.9\textwidth}
\captionsetup{format=plain, labelfont=bf}
\caption{
The left graphic shows $H$ as a function of $(\phi,\dot\phi)$ in terms of its level sets as it is obtained by locally solving the solution set of the energy equation \eqref{eq:energy_equation_in_phase_space_section} for $(\phi,\dot\phi)$ around $\big(0,0,H_{\textup{vac}}(\Lambda)\big)$. Exemplary, we have chosen the parameters $\Lambda=m=1$, $\xi=\frac1{12}$. The light shaded area marks points $(\phi,\dot\phi)$ where of $0=f(\phi,\dot\phi,H)$ possesses no real solution for $H$, where the dark shaded area marks points $(\phi,\dot\phi)$ where real solutions exist, but these are negative and hence disconnected from the part of the solution surface presented here. The right graphic shows both sheets of the solution surface in a three-dimensional view. The ellipses mark constant-$H$-levels as in (a), but equidistantly with distance $\frac12$.} \label{fig:level_set_plots}

\end{minipage}
\end{figure}

For a rough overview, we present some characteristic plots in phase space $(\phi,\dot{\phi})$. Those are sections of the surface in $(\phi,\dot{\phi},H)$ studied, for some constant $H$. 

Figure~\ref{fig:level_set_plots} (a) is a graphic of the surface of solutions to the energy equation \eqref{eq:energy_equation_in_phase_space_section} for $\Lambda=m=1$ and $\xi=\frac{1}{12}$.
Notice that since $f\big(0,0,H_{\textup{vac}}(\Lambda)\big)=0$ and $\nabla_{(\phi,\dot\phi,H)}f\big(0,0,H_{\textup{vac}}(\Lambda)\big)=\big(0,0,-12H_{\textup{vac}}(\Lambda)\big)\neq0$, it is indeed possible, locally around $(\phi,\dot\phi,H)=\big(0,0,H_{\textup{vac}}(\Lambda)\big)\approx(0,0,0.577)$, to represent this surface as a graph of a function of $\phi$ and $\dot\phi$ of which the level sets are shown. The shown domain for this $H(\phi,\dot\phi)$ in Figure~\ref{fig:level_set_plots} (a) (i.e.\ the unshaded area) is bounded both from the left and the right by points where this representation cannot be continued any further. 

One can see how the level sets are given by arcs of ellipses. Indeed, for the presently discussed parameters we have $\det(A_1)=1+H^2/4>0$ and $\tr(A_1)=2+H^2/2>0$ (cf.\ \eqref{eq:det_and_trace_for_conic_section_matrix}), so $A_1$ is positive definite for all $H$. Consequently, the conic section solving \eqref{eq:energy_equation_in_phase_space_section} at fixed $H$ is empty if $2\Lambda-6H^2<0$, a single point if $2\Lambda-6H^2=0$ and an ellipse if $2\Lambda-6H^2>0$. Note that the graphic may be extended to negative $\dot\phi$ values by a reflection about the origin $(\phi,\dot\phi)=(0,0)$. Also note that missing parts of the ellipses on the left boundary of the domain belong to another part of the surface which lies above (at larger $H$-values) the shown part. The complete ellipses are shown as conic sections of the energy equation surface in $(\phi, \dot{\phi},H)$ space in Figure~\ref{fig:level_set_plots} (b). 

If one explicitly writes out the roots of the second order polynomial $H\mapsto f(\phi,\dot\phi,H)$, it is a simple task to verify that if these roots are real (corresponding to the complement of the light shaded area in Figure~\ref{fig:level_set_plots} (a)), then they are both positive for $\phi<-2\sqrt{3}$ (we show in Figure~\ref{fig:level_set_plots} the smaller one), one positive and one negative for $-2\sqrt{3}<\phi<2\sqrt{3}$ (we show the positive one in Figure~\ref{fig:level_set_plots} (a)) and both negative for $\phi>2\sqrt{3}$ (corresponding to the darker shaded area in Figure~\ref{fig:level_set_plots} (a)).

\subsubsection[Singular points for \texorpdfstring{${m=\Lambda=0}$}{sth}]{Singular points for \texorpdfstring{$\mathbf{m=\Lambda=0}$}{sth}}
\label{sec:singular_points}
We briefly comment on the singular points of the energy equation's solution surface at $H=0$ in the case $\Lambda=0$ concentrating on $0<\xi<\frac16$. However, for other $\xi$-values one can apply similar arguments.

If we set $m=0$, we noted above that the blow-up substituted energy equation \eqref{eq:other_blowup_substituted_energy_conic_sections} is a conic section equation for $\phi$ and $\dot\phi$ which is independent of $H$. The determinant and trace of the relevant part of the matrix representation of \eqref{eq:other_blowup_substituted_energy_conic_sections} read
\begin{equation}
\label{eq:det_and_tr_blow-up_substituted}
\det\begin{pmatrix}
    1&6\xi\\6\xi&6\xi
\end{pmatrix}
=6\xi(1-6\xi)
\qquad\textup{and}\qquad
\mathrm{tr}\begin{pmatrix}
    1&6\xi\\6\xi&6\xi
\end{pmatrix}=1+6\xi\,.
\end{equation}
Thus, for $0<\xi<\frac16$ we have a certain ellipse at each $H$ which, by our discussion above, touches both the $\phi=\pm\phi_{\textup{crit}}(\xi)$-hyperplanes, namely in the points of the graphs of \eqref{eq:parametrizations_of_lines_at_phi_crit} (under the same blow-up). Reversing the blow-up substitution results in a family of ellipses parametrized by $H$ whose semi-major axes' lengths converges to $\phi_{\textup{crit}}(\xi)$ as $H\to \infty$ and whose semi-minor axis' length converges to 0. Since the blow-up substitution in consideration is of order 1, the latter limit is approached with a linear leading order. Consequently, any point 
\begin{equation}
\label{eq:singular_points_at_H_equals_0_with_zeromass}
    \big\{(\phi,\dot\phi,H)=(\phi,0,0)\big| -\phi_{\textup{crit}}(\xi)<\phi<\phi_{\textup{crit}}(\xi)\big\}
\end{equation} 
belongs to the surface and representing the (non-negative-$H$-part of the) surface as a function $(\phi,\dot\phi)\mapsto H(\phi,\dot\phi)$ yields that
\begin{equation}
    H(\phi,\dot\phi)\in\alpha_\phi|\dot\phi|+\mathcal{O}(\dot\phi^2)\,
\end{equation}
as $\dot\phi\to0$, for any fixed $\phi\in\big(-\phi_{\textup{crit}}(\xi),\phi_{\textup{crit}}(\xi))$ and some $\alpha_\phi>0$. Consequently, any point in \eqref{eq:singular_points_at_H_equals_0_with_zeromass} is a singular point of the surface.

If we assume $m>0$ instead, we can apply another blow-up and study
\begin{equation}
\label{eq:double_blowup_energy_equation}
    0=\frac1{H^2}f(H\phi,H\dot\phi,H)=
    \begin{pmatrix}
	\dot\phi&\phi&1
	\end{pmatrix}
    \begin{pmatrix}
	1&6\xi H&0\\
	6\xi H&m^2+6\xi H^2&0\\
	0&0&-6
	\end{pmatrix}
	\begin{pmatrix}
	\dot\phi\\\phi\\1
	\end{pmatrix}\,.
\end{equation}
The matrix coincides with the matrix in \eqref{eq:blowup_substituted_energy_conic_sections}, up to the bottom right entry that here does not depend on $H$ anymore. For $0<\xi<\frac16$ we have two positive eigenvalues and the bottom right entry being negative implies the solution sets of \eqref{eq:double_blowup_energy_equation} around $H=0$ is some family of ellipses that vary continuously with $H$, bounded away from the point $(H\phi,H\dot\phi,H)=(0,0,0)$. Particularly, the ellipse at $H=0$ fulfills the equation
\begin{equation}
    m^2\phi^2+\dot\phi^2=6
\end{equation}
and thus has a semi-major axis of length $\sqrt{6}$ parallel to the $\dot\phi$-direction and a semi-minor axis of length $\frac{\sqrt{6}}{m}$ parallel to the $\phi$-direction. We reverse the blow-up and find that the (non-negative-$H$-part of the) surface, represented as a function $(\phi,\dot\phi)\mapsto H(\phi,\dot\phi)$, fulfills
\begin{equation}
H(\phi,\dot\phi)\in\sqrt{6}~\big\|\big(\tfrac1m\phi,\dot\phi\big)\big\|+\mathcal{O}(\|(\phi,\dot\phi)\|^2)
\end{equation}
as $\|(\phi,\dot\phi)\|\to0$.
Hence, the point $(\phi,\dot\phi,H)=(0,0,0)$ is indeed a singular point. 

\subsection{Constraints from a prescribed initial deceleration parameter}
\label{sec:deceleration_constraints}

In order to simulate solutions to our model of classical Klein-Gordon cosmology, we have to set initial values at $t=0$. As we have seen in the previous section, this is a non-trivial task as for certain choices of parameters $m$, $\xi$ and $\Lambda$ and initial values $a(0)$ and $\dot a(0)$ real values for $\phi(0)$ and $\dot\phi(0)$ that solve the energy equation \eqref{eq:energy_equation_in_phase_space_section} do not necessarily exist. In the present section we want to fix all four of the initial values. As we mentioned earlier, $a(0)$ and $\dot{a}(0)$ can, without loss of generality, be set to 1. In the previous section we studied the constraints of the energy equation which can be used to fix one of the $\phi(0)$ or $\dot{\phi}(0)$ (in suitable bounds). A fixed deceleration parameter $q_0=q(0)$ fixes the last initial condition. As we will see, the requirement for $\phi(0)$ and $\dot{\phi}(0)$ to be real for each $q_0$ constrains the parameter space.

We consider the trace equation in the form
\begin{equation}
\label{eq:trace_equation_with_q}
0
=
H^2(q-1)+H^2(q-1)\xi(6\xi-1)\phi^2
+\frac{2\Lambda}{3}+\frac{6\xi-1}{6}
\dot{\phi}^2
+\frac{(2-6\xi )m^2}{6}\phi^2\,,
\end{equation}
which is particularly an algebraic equation for $a$, $\dot a$, $\phi$, $\dot\phi$ and $q=q[\,a\,]=-\frac{a\ddot a}{\dot a^2}$ at each time.
We denote $H=\frac{\dot a}{a}$ as before and since we are mostly interested in this equation at $t=0$ we leave out any arguments of the dynamic variables and of $q$.

Supposed we have found initial values for $a$, $\dot a$, $\phi$ and $\dot\phi$ that solve the energy equation \eqref{eq:energy_equation_in_phase_space_section}, then \eqref{eq:trace_equation_with_q} can be solved for $q$ as
\begin{equation}
\label{eq:q_of_phi_dphi_H}
q =1-\frac{4\Lambda+  (2-6\xi)m^2\phi^2+(6\xi-1)\dot{\phi}^2}{6H^2\big(1+ \xi(6\xi-1)\phi^2\big)} \,,
\end{equation}
provided that $H\neq0$ and $|\phi|\neq\phi_{\textup{crit}}(\xi)$ if $0<\xi<\frac16$ (for instance, choose $\Lambda>0$ or $m>0$, cf.\ \eqref{eq:confinement_set_for_solution_set} and the remark thereafter). These values, however, might become quite large as we will see in the following example.

In a straight-forward computation one finds that $f$ vanishes along the curve
\begin{equation}
\label{eq:curve_for_divergent_q}
\Big(-\frac{1}{\sqrt{24}},-\frac{1}{\sqrt{12}}\Big)\to\mathbb R^3\,,~\phi\mapsto\left(\phi,-\sqrt{\frac{2 (\phi^2-12) (2 + \phi^2)}{24-\phi^2}},\sqrt{\frac{4\phi^2 + 2\phi^4}{(\phi^2-12)(24-\phi^2)}}\right)\,,
\end{equation}
for the exemplary parameters $\Lambda=m=1$, $\xi=\frac1{12}$ (as in Figure~\ref{fig:level_set_plots}) and that \eqref{eq:q_of_phi_dphi_H} evaluated along this curve yields
\be
q=\frac{2880 + 816\phi^2 - 122\phi^4 + 4\phi^6}{\phi^2(24-\phi^2)(\phi^2+2)}\,.
\ee
While the numerator polynomial has no real root, $\phi_{\textup{crit}}(\frac{1}{12})=\frac{1}{\sqrt{24}}$. In particular, $q$ diverges towards the left bound of the domain interval in \eqref{eq:curve_for_divergent_q}.

On the other hand, by the discussion in Section~\ref{sec:constantEOS_and_LCDM_cosmology} we have a rough idea of a preferable magnitude for an initial value $q_0$ of $q$ in or at least close to the interval $[-1,1]$. In turn, we have no such expectation for $\phi$ or $\dot\phi$. Thus, a preferable situation would be to have control over the  $q_0$, view that as a parameter of our model and get initial values for $\phi$ and $\dot\phi$ out from our model. This results in four, two or zero solutions for $(\phi,\dot\phi)$-pairs. However, the pairs of solutions are always degenerate, resulting at most to two inequivalent cosmologies by our models symmetry under $\phi\mapsto-\phi$. 

We rewrite \eqref{eq:trace_equation_with_q} (multiplied by 6) for fixed $q$ and $H$ into a conic section equation in $(\dot\phi,\phi,1)$ with the matrix
\begin{equation}
\label{eq:trace_equation_with_q_matrix_representation}
	B
	=
	\mathrm{diag}\left(
		6\xi-1
		~,~
		H^2(q-1)6\xi(6\xi-1)+(2-6\xi )m^2
		~,~
		6H^2(q-1)+4\Lambda
	\right)
\end{equation}
and want to study the intersection of the solution conic section with that of the energy equation \eqref{eq:energy_equation_in_phase_space_section}. 

\begin{figure}
\centering
\begin{subfigure}{.32\textwidth}
    \centering
    \includegraphics{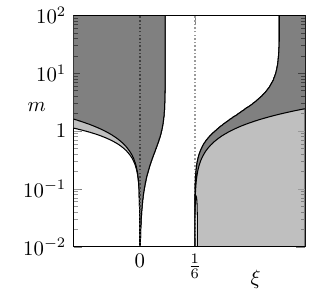}
    \caption{$\Lambda=0$, $q_0=q_{\Lambda\textup{CDM}}$}
\end{subfigure}
\hfill
\begin{subfigure}{.32\textwidth}
    \centering
    \includegraphics{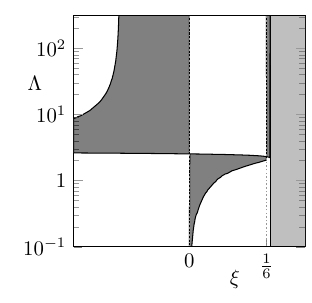}
    \caption{$m=0$, $q_0=q_{\Lambda\textup{CDM}}$}
\end{subfigure}
\hfill
\begin{subfigure}{.32\textwidth}
    \centering
    \includegraphics{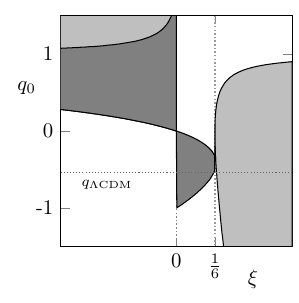}
    \caption{$\Lambda=2$, $m=0$}
\end{subfigure}

\smallskip
\begin{minipage}{.9\textwidth}
\captionsetup{format=plain, labelfont=bf,singlelinecheck=false}
\caption{The graphics depict parameter space sections where there exist real initial values for $\phi$ and $\dot\phi$ as solutions for \eqref{eq:energy_equation_in_phase_space_section} and \eqref{eq:trace_equation_with_q} with fixed $q$. The lighter gray areas mark parameters where there are merely two solutions that are related by $(\phi_0,\dot\phi_0)\leftrightarrow(-\phi_0,-\dot\phi_0)$. The darker gray areas mark parameters in which there exist four solutions, that is, parameters for which two inequivalent cosmologies are possible. Throughout all graphics we have $a(0)=1$, $\dot a(0)=1$ and $\kappa=1$ as argued in the text. The remaining parameters $\Lambda$, $m$, $\xi$ and $q_0$ are chosen as indicated in the subfigures' captions and the axes' labels.
\label{fig:initial}
}
\end{minipage}
\end{figure}

In general, two centered, non-degenerate conic sections are either equal or they intersect in four points at most. Therefore, we can solve either the energy \eqref{eq:energy_equation_in_phase_space_section} or the trace equation \eqref{eq:trace_equation_with_q} either for $\phi$ or for $\dot\phi$, replace the result into the respective other equation and therefrom derive a fourth-order polynomial equation for the respective other variable. Fourth-order polynomials have a closed solution formula and we obtain four $(\phi,\dot\phi)$-pairs which may be real or not, and non-degenerateness of the conic sections implies that one way or the other, this algorithm goes through. These solutions can, for instance, be determined by computer algebra software and we omit to give lengthy formulas here. Notice that the origin $(\phi,\dot\phi)=(0,0)$ cannot be contained in a non-degenerate conic section. In particular, in the case of non-degenerate conic sections our algorithm of determining initial values cannot result in a vacuum solution with $\phi(t)=0$ for all $t$  and, since centered conic section are symmetric under $(\phi,\dot\phi)\mapsto(-\phi,-\dot\phi)$, we always obtain an even number of solutions, two of which are related under said symmetry and yield the same cosmology (i.e.\ the same scaling factor $a$).  We want to forgo a detailed discussion on the cases when these conic sections are degenerate, merely we note that we have come across such a case in Section~\ref{sec:Special_values_of_the_parameters}, where we were free to choose the parameter $\phi(0)$ in \eqref{eq:special_case_confcoup_massless_Lambda0} without affecting the initial deceleration parameter. We emphasize that in this setting the resulting cosmology  did not depend on this parameter. 

\begin{figure}
\centering
    \includegraphics{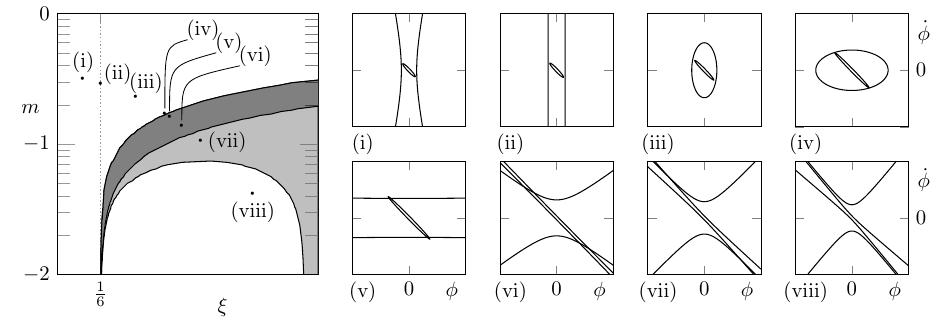}
\begin{minipage}{.56\textwidth}
\captionsetup{format=plain, labelfont=bf,singlelinecheck=false}
\caption{
The left graphic shows a zoom into Figure~\ref{fig:initial} (a) close to $\xi=\frac16$, namely for $0.165<\xi<0.175$. All three possibilities of four, two or no solutions are present in these parameter bounds. The right eight graphics show the associated conic sections in the $(\phi,\dot\phi)$-plane solving the energy equation \eqref{eq:energy_equation_in_phase_space_section} and the trace equation \eqref{eq:trace_equation_with_q} for $-70<\phi,\dot\phi<70$, with specific parameter choices as in the tabular on the right and indicated in the left graphic. Recall that in Figure~\ref{fig:initial} (a), and hence also here, we fixed $\Lambda=0$, $q_0=q_{\Lambda\textup{CDM}}$, $a(0)=\dot a(0)=H=1$. In each graphic, the trace equation's conic section can uniquely be recognized as the one whose symmetry axes are parallel with the $\phi$- and $\dot\phi$-axes.
\label{fig:initial_zoomed}
}
\end{minipage}
\quad
\begin{minipage}{.39\textwidth}
    \includegraphics{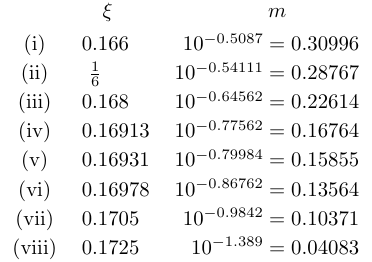}

    ~
\end{minipage}
\end{figure}

As described above, given initial values $a(0)$ and $\dot a(0)$ and parameters $\Lambda$, $m$, $\xi$ and $q_0$ as the prescribed initial deceleration parameter, we obtain (generically) up to four $(\phi,\dot\phi)$-pairs as possible initial values for our model.
We show in Figure~\ref{fig:initial}, exemplary, three graphics with fixed $a(0)=\dot a(0)=1$ (i.e.\ $H=1$). From the four free parameters we fix two and allow the other two to vary. The coupling constant $\xi$ is always a free parameter. Each point is marked in white, light gray or dark gray, according to the existence of, respectively, no, two or four solutions of both \eqref{eq:energy_equation_in_phase_space_section} and \eqref{eq:trace_equation_with_q} for $(\phi,\dot\phi)$. 

Moreover, we show in Figure~\ref{fig:initial_zoomed} a zoom into Figure~\ref{fig:initial} (a). On the one hand, we want to show that what in Figure~\ref{fig:initial} (a) looks as a numeric artefact at first sight, indeed corresponds to an area where no solutions exist. On the other hand, around these parameters the conic sections associated to \eqref{eq:energy_equation_in_phase_space_section} and \eqref{eq:trace_equation_with_q} show a rich variety of different behavior in terms of degeneracy, their shapes and how they intersect or not. These are schematically depicted also in Figure~\ref{fig:initial_zoomed}. Notice that since the representation matrix for the trace equation in \eqref{eq:trace_equation_with_q_matrix_representation} is diagonal, its eigenvectors are parallel with the $\phi$- and with the $\dot\phi$-axis, respectively. We did not label the two conic sections as the one associated with the trace equation can be unambiguously identified as it has a symmetry under vertical and horizontal reflections. Different situations arise based on the relative positions of the conic sections and how they intersect or not. 

For instance, the upper boundary of the dark gray area around the parameter point (iv) marks points where the two ellipses are tuned to touch in two points. Hence, the set of allowed parameters is closed around this point. As another example, we can see how the lower bound of the light gray area in between the points (vii) and (viii) marks points where the two hyperbolas share an asymptote. Approaching such a point from above, the two hyperbolas intersect in two points that tend to infinity and in the limit. When the asymptotes eventually coincide, there is no more common point. Hence, the set of allowed parameters is open around this boundary. A similar effect happens in between the points (vi) and (vii) when we pass from a four-solution regime into a two-solutions regime.

\subsection{Reparametrization of the dynamical system}
In order to study singularities of a dynamical system numerically, it is sometimes helpful to perform a substitution in the underlying ODE. Therefore, we want to express the dynamical system defined by \eqref{eqn:syst} in terms of $(\phi,\dot\phi,H)$ as parametrized by $a$. Then, the number of dynamic variables is reduced by one and we have a numerical tool to study and confirm the singular behavior of solutions using a second independent method.

More precisely, under the assumption that $\dot a$ is positive, the scaling factor $a$ is an invertible function of time. Hence, the ODE for the dynamical variables $(a,\dot a,\phi,\dot\phi)$ defined by \eqref{eqn:syst} imposes an ODE for the dynamical variables
\begin{equation}
\mathcal{H}:=\left(\frac{\dot a}{a}\right)\circ a^{-1}\,,
\quad
\Phi:=\phi\circ a^{-1}
\quad\textup{and}\quad
\Pi:=\dot\phi\circ a^{-1}\,.
\end{equation}
We denote the independent variable of this ODE as $a$ again. The vector fields imposed by \eqref{eqn:syst} read as
\blea
\frac{\mathrm{d}\mathcal{H}}{\mathrm{d}a}
 &=&-2\frac{\mathcal{H}}{a}
+\frac{1}{6a\mathcal{H}}\frac{4\Lambda
+(6\xi-1)\Pi^2
+2(1-3\xi)m^2\Phi^2
}{1+\xi(6\xi-1)\Phi^2}\,,
\\[2pt]
\frac{\mathrm{d}\Phi}{\mathrm{d}a}&=&\frac{\Pi}{a\mathcal{H}}\,,
\\[2pt]
    \frac{\mathrm{d}\Pi}{\mathrm{d}a}&=&-3\frac{\Pi}{a}-\frac\Phi{a\mathcal{H}}\,\frac{4\Lambda \xi+m^2
+\xi(6\xi-1)\Pi^2
+\xi m^2\Phi^2
}{1+\xi(6\xi-1)\Phi^2}\,,
\elea
where the energy equation is
\begin{equation}
0=f\big(\Phi(a),\Pi(a),\mathcal{H}(a)\big) \,,
\end{equation}
for all $a$, with the function $f$ as defined in \eqref{eq:energy_equation_in_phase_space_section}. This reparametrization works on any interval on which $\dot a$ is positive. However, we note that $\mathcal{H}>0$ along a solution $a$ already implies $\dot a$ to be positive. Moreover, we recall that whenever $\Lambda>0$, the energy possesses no solution with $H=0$, wherefore any solution with $\dot a(0)>0$ ($\dot a(0)<0$) is strictly increasing (decreasing) and invertible, allowing for the substitution above.

In any numerical solution in the following it is cross-checked whether the solution in the above reparametrized dynamical system shows the same properties. Whenever a comparison is possible we observe that solutions and functions derived from a solution (such as $H[\,a\,]$/$\mathcal{H}$ or trajectories $\big(\phi(\cdot),\dot\phi(\cdot)\big)$/$\big(\Phi(\cdot),\Pi(\cdot)\big)$) are (numerically) indistinguishable between the two parametrizations employed. We remark that by computing
\be
\frac{\mathrm{d}}{\mathrm{d}t} H[\,a\,](t)=-H[\,a\,](t) \big(q[\,a\,](t)+1\big)\,,
\ee
we find
\begin{equation}
    \mathcal{Q}:=q[\,a\,]\circ a^{-1}=-\frac{1}{\mathcal{H}}\frac{\mathrm{d}\mathcal{H}}{\mathrm{d}a}-1 \,,
\end{equation}
in order to compare trajectories involving $q$.

\section{Evolution of the scale factor}
\label{sec:evolution}

In this section, we want to study solutions for $a(t)$ explicit for different values of the parameters by a numerical approach. Here, we are especially interested in the early and late time behavior of the scale factor. In particular, the first two subsections are discussing the two types of singularities introduced in Section~\ref{sec:Types_of_cosmological_singularities}. The third subsection is about the late-time behavior of the scale factor. 

We will not discuss solutions with $\xi<0$ throughout this section as negative coupling usually results in unhysical systems. 

\subsection[\texorpdfstring{$\gamma$}{sth}-type solutions]{\texorpdfstring{$\mathbf\gamma$}{sth}-type solutions}

Throughout the present section we refer to $\gamma$-type solutions as those for which asymptotically $a(t)\to 0$ and $\Gamma[a](t) \to \gamma$ as $t \to t^s$, for $\gamma\in\mathbb R$. So while $\Gamma$ is not necessarily constant at all times, at early times, it asymptotically approaches one of the solutions of $\gamma$-type stress-energy tensors discussed in Section~\ref{sec:constantEOS_and_LCDM_cosmology}. These solutions appear for $\xi>1/6$ while we fix $\Lambda$ and $m$ and vary $q_0$. 

\subsubsection{Singular behavior}

In terms of their singular behavior we recognize three classes of $\gamma$-type solutions: (I) The pure radiation expansion (II) The $\gamma$-type Big Bang singularity (defined in Definition~\ref{def:bigbang}) and (III) The scale factor asymptotically approaches zero. 

\begin{figure}
\centering
\begin{subfigure}[b]{.47\textwidth}
    \includegraphics{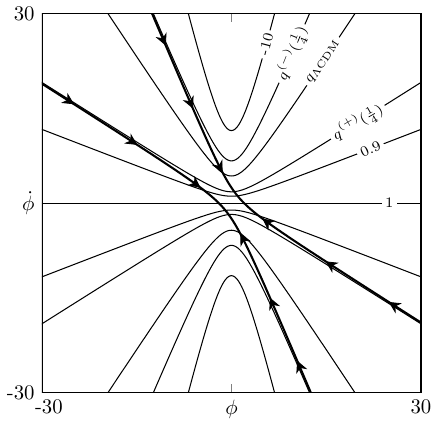}
\end{subfigure}
    \hfill
\begin{minipage}[b]{.49\textwidth}
\captionsetup{format=plain, labelfont=bf,singlelinecheck=false}
\caption{
The graphic shows a family of conic sections for consistent initial values in the parameter setting $m=\Lambda=0$, $\xi=\frac14$. The thick hyperbola corresponds to the the energy equation \eqref{eq:energy_equation_in_phase_space_section}, the thin hyperbolas correspond to the trace equation \eqref{eq:trace_equation_with_q}, where the line labels indicate the respective value of $q_0$. The arrows on the energy equations solution hyperbola indicate the the direction of the time flow under the dynamics of our model, that is, $\phi$ is decreasing if $\dot\phi<0$ (arrows point left-ward) and increasing if $\dot\phi>0$ (arrows point right-ward).
\label{fig:parameter_survey_conic_sections}
}
\end{minipage}
\end{figure}

As a first interesting parameter setting we want to discuss a massless field with curvature coupling $\xi=\frac14$ without a cosmological constant, $\Lambda=0$. We will discuss below how this particular $\xi$-value is to some extend generic for $\xi>\frac16$ (or say for $\xi>\frac{3}{16}$, respectively). As usual, we assume $a(0)=\dot a(0)=1$ and hence $H=1$ at $t=0$. This case includes all three classes of solutions depending on the value of $q_0$. The conic sections as the solution sets of the energy equation \eqref{eq:energy_equation_in_phase_space_section} and the trace equation \eqref{eq:trace_equation_with_q} with various values for $q_0$ are shown in Figure~\ref{fig:parameter_survey_conic_sections}. We define the two values 
\begin{equation}
\label{eq:qplusminus}
q^{(\pm)}(\tfrac14)=-1\pm\sqrt{3}
\qquad\textup{with}\qquad 
q^{(+)}(\tfrac14)\approx0.732
\qquad\textup{and}\qquad 
q^{(-)}(\tfrac14)\approx-2.732
\end{equation}
which are obtained by claiming that an asymptote of the trace equation hyperbola coincides with an asymptote of the energy equation hyperbola. We have four initial value pairs for $(\phi,\dot\phi)$ for $q_0\in\big(q^{(+)}(\tfrac14),1\big)$, two initial value pairs for $q_0\in\big(q^{(-)}(\tfrac14),q^{(+)}(\tfrac14)\big]$ or $q_0=1$ and no initial value pairs for the remaining $q_0$ values. We note that these values $q^{(\pm)}(\tfrac14)$ do not depend on a choice of initial $H$. Therefore recall that the energy equation's conic section for our presently chosen parameters ($m=\Lambda=0$) are independent of $H$ in the blow-up coordinates $(\phi,H\dot\phi)$ and note that the same is true for the trace equation's conic section. In particular, for different $H$-values the situation in Figure~\ref{fig:parameter_survey_conic_sections} is merely stretched in the vertical direction.

The first class of solutions (I) is given by pure radiation expansion
\begin{equation}
\label{eq:pure_radiation_expansion_solution_with_constant_field}
a(t)\propto\sqrt{1+2t}
\qquad\textup{and}\qquad
\phi(t)=\pm2\
\end{equation}
for all $t$ (and time-translations thereof), as we have determined in Section~\ref{sec:Special_values_of_the_parameters} (cf.\ \eqref{eq:radiation_solutions_when_m_equals_lambda_equals_0}).
In fact, this is the only (non-Minkowski) solution type for $m=\Lambda=0$ which can ever attain $\dot\phi(t)=0$ at an instance of time $t$. Therefore note that $\dot\phi(t)=0$ and $H[\,a\,](t)\neq 0$ at some $t$ implies $q[\,a\,](t)=1$ via \eqref{eq:q_of_phi_dphi_H} as well as $\phi(t)\in\{\pm2\}$ via \eqref{eq:energy_equation_in_phase_space_section}, hence by translation and scaling we can find a solution of the type \eqref{eq:pure_radiation_expansion_solution_with_constant_field} that matches in all initial values. Assuming, on the other hand, $\dot\phi(t)=0$ and $H[\,a\,](t)=0$ at some $t$, there exists a Minkowski solution matching all initial values in $t$ with a constant field value as we have found in Section~\ref{sec:Minkowski_and_De_Sitter_solutions}. 

\begin{figure}
\centering
    \includegraphics{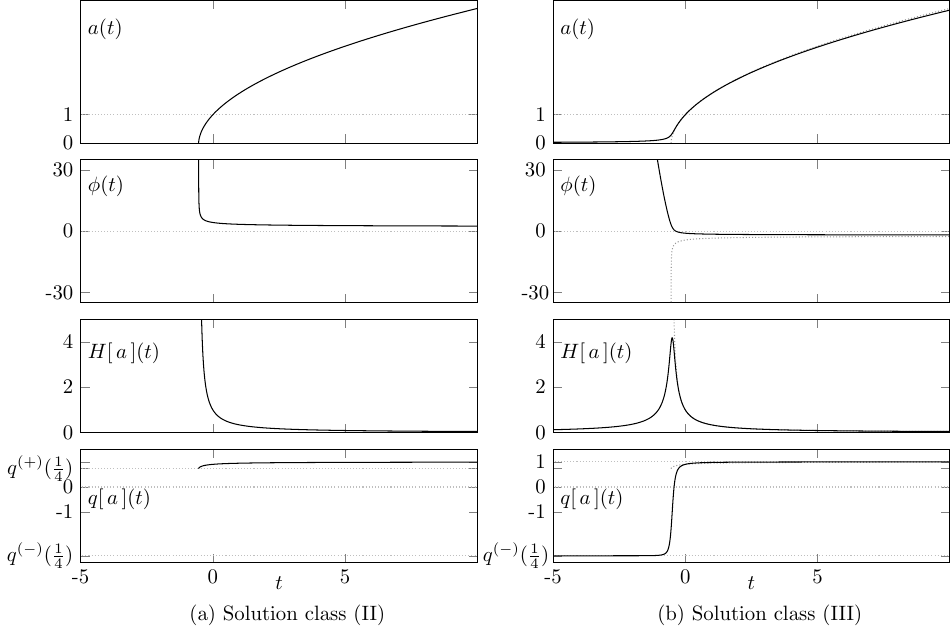}
    \qquad
\begin{minipage}{.9\textwidth}
\captionsetup{format=plain, labelfont=bf,singlelinecheck=false}
\caption{
The graphic shows the two inequivalent cosmologies of the classes (II) and (III) for the parameter setting $\Lambda=m=0$ and $\xi=\frac14$ in terms of the scaling factor $a$, the field $\phi$, the Hubble rate $H[\,a\,]$ and the deceleration parameter $q[\,a\,]$. We have chosen initial values $a(0)=\dot a(0)=1$ and $q_0=0.9$, hence we have a total of four initial value pairs for $(\phi,\dot\phi)$. The remaining two solutions are obtained from the shown ones by the symmetry $\phi(t)\mapsto-\phi(t)$ for all $t$. For a rough comparison we have included in the right plots as dotted gray lines the curves of the left plots (with a "$-$"-sign for $\phi$ in order to match the late-time limits).
\label{fig:parameter_survey_first_solution}
}
\end{minipage}
\end{figure}

The second class of solutions (II) is shown in Figure~\ref{fig:parameter_survey_first_solution}(a). At early times we see a Big Bang singularity in which $\phi$ (and thus also $\dot\phi$) and $H[\,a\,]$ diverge and $q[\,a\,]$ approaches the value $q^{(+)}(\tfrac14)$. We note that for this solution we have a positive, decreasing $\phi$ at initial time, that is, an initial value for $(\phi,\dot\phi)$ from the lower right quadrant in Figure~\ref{fig:parameter_survey_conic_sections} (or, equivalently, the upper left quadrant). Above we noted that the conic sections in Figure~\ref{fig:parameter_survey_conic_sections} only depend on $q_0$ if we express them in the $\big(\phi,H\phi\big)$-plane. In particular, we can see that as $q_0$ decreases, the conic sections' intersection in the lower right quadrant moves towards larger $\phi$-values until, eventually as $q_0\to q^{(+)}(\tfrac14)$ from above, this intersection point ceases to exist. Hence, the presently discussed solution class does not allow for $q$-values outside $\big(q^{(+)}(\tfrac14),1\big)$ and indeed, numerically we observe a monotonic $q[\,a\,]$ that ranges through this interval.

The last class of solutions (III) is shown in Figure~\ref{fig:parameter_survey_first_solution} (b). The main difference to the previous solution class is that now initial values for $(\phi,\dot\phi)$ are contained in the lower left (or upper right) quadrant in Figure~\ref{fig:parameter_survey_conic_sections}, that is, we have an initially negative, decreasing (or positive, increasing) field. In this third solution class we observe a quite similar, radiation-like late-time behavior. Also here, the solution appears to be well approximated by \eqref{eq:pure_radiation_expansion_solution_with_constant_field} at late times. However, the early time behavior is very different. In fact, we observe that the scale factor asymptotically approaches zero and not a Big Bang singularity. Therefore, we observe in Figure~\ref{fig:parameter_survey_first_solution} that $q[\,a\,]$ tends towards $q^{(-)}(\tfrac14)$, corresponding via \eqref{eq:deceleration_state_fraction_correspondence} to a EOS coefficient smaller than $-1$ (the latter corresponding to typical Dark Energy in terms of a positive $\Lambda$). This effect is discussed more in Section~\ref{sec:inflation}. The transition from the ``$q[\,a\,]\!=\!q^{(-)}(\tfrac14)$''- to the ``$q[\,a\,]\!=\!1$''-regime happens in a quite steep manner, and the corresponding peak in $H[\,a\,]$  appears surprisingly symmetric (but is, in fact, not). 

Notice that, in regard of the other two solutions obtained by $\phi(t)\mapsto-\phi(t)$ for all $t$ (with initial values in the upper two quadrants of Figure~\ref{fig:parameter_survey_conic_sections}), the solution classes are better distinguished by which branch of the energy equation's hyperbola the initial values $(\phi,\dot\phi)$ lie on. Again, if we express the latter hyperbola in terms the coordinates $(\phi,H\dot\phi)$, it becomes independent of the dynamics. In other words, for any solution $t\mapsto\big(a(t),\phi(t)\big)$ the map
\begin{equation}
    \label{eq:rescaled_hyperbola_representation}
t\mapsto\left(\phi(t),\frac{\dot\phi(t)}{H[\,a\,](t)}\right)
\end{equation}
takes values only in the aforementioned hyperbola (away from $H[\,a\,]\neq 0$). Since a solution cannot pass through the $\dot\phi=0$-axis, this hyperbola is divided into four branches that must be invariant under the dynamics, two of which allow for sign changes in $\phi$ and two do not. As we can see in Figure~\ref{fig:parameter_survey_conic_sections}, the two branches that do not allow for sing changes in $\phi$ allow only for $q$-values in $\big(q^{(+)}(\tfrac14),1\big)$, whereas the other two branches allow $q$-values in $\big(q^{(-)}(\tfrac14),1\big)$. Checking for the signs of $\dot\phi$ along the branches, we can conclude that under the dynamics, the pairs \eqref{eq:rescaled_hyperbola_representation} run along the branches, always towards the points $(\phi,\dot\phi)=(\pm2,0)$, as indicated by the arrows in Figure~\ref{fig:parameter_survey_conic_sections}.

We can  also read off the same behavior from the matrix representations for the energy and the trace equation in Section~\ref{sec:Parameter_constraints}. Whenever $\xi>\frac{1}{6}$ and $m=0$, then $\det{A}$ and $\det{A_1}$ are negative and thus the energy equation is always solved by a hyperbola in the $(\phi,\dot\phi)$-plane. The trace equation (with $\Lambda=0$) is solved by a hyperbola for $q<1$, the line defined by $\phi=0$ (a degenerate hyperbola) for $q=1$ and has no solutions if $q>1$. Claiming that either asymptote of the energy equation's hyperbola coincides with either asymptote of the trace equation's hyperbola two functions 
\begin{equation}
q^{(\pm)}:(\tfrac{1}{6},\infty)\to(-\infty,1)
\end{equation}
which generalize $q^{(\pm)}(\frac14)$ from above to arbitrary values of $\xi$. A plot of these functions is shown in Figure~\ref{fig:parameter_survey_first_solution_variations} (a). Our above argumentation works for any $\xi>\frac16$ and the respective values of the functions $q^{(\pm)}(\cdot)$ determine the early-time behavior of the emerging solution classes.

The class of solutions (I) in\eqref{eq:pure_radiation_expansion_solution_with_constant_field} exists independently of $\xi$ and is again a degeneration of the other two classes. The second class (II) shows a monotonic transition from a Big Bang with $q^{(+)}(\xi)$-type matter (in the correspondence from \eqref{eq:deceleration_state_fraction_correspondence}) to a radiation-type late-time expansion, whereas the third class (III) is a monotonic transition from $q^{(-)}(\xi)$-type matter at early times to radiation at late times.

\subsubsection{Inflation}
\label{sec:inflation}

\begin{figure}
\centering
    \includegraphics{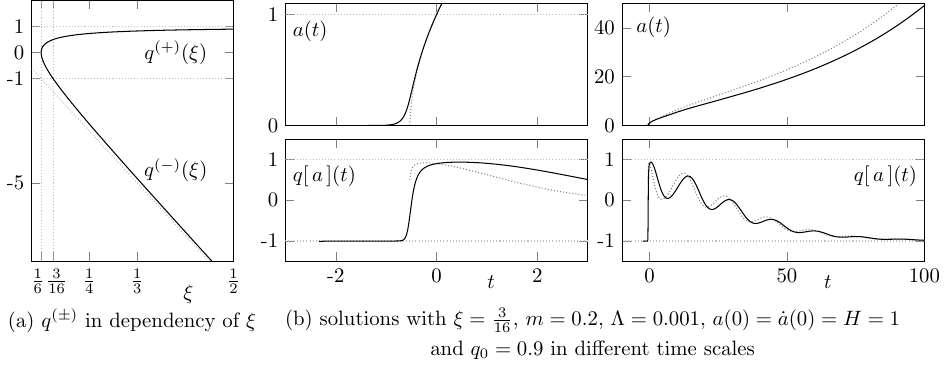}

\smallskip
\begin{minipage}[b]{.9\textwidth}
\captionsetup{format=plain, labelfont=bf,singlelinecheck=false}
\caption{
The graphic in (a) shows plot of the two $q$-values that distinguish the different parameter regimes in dependency of $\xi$, as for example in \eqref{eq:qplusminus} at $\xi=\frac{1}{4}$. The asymptotic expansions as $\xi\to\infty$, that is $q^{(+)}(\xi)\in 1+\mathcal{O}(\xi^{-1})$ and $q^{(-)}(\xi)\in -24\xi+3+\mathcal{O}(\xi^{-1})$, are shown as dotted gray lines. Also the value $-1$ corresponding to traditional Dark Energy via \eqref{eq:deceleration_state_fraction_correspondence}, is marked as such a line. (b) shows the two inequivalent solutions to the given parameters, each in two different time scales. Particularly, we have chosen $\xi=\frac3{16}$ to obtain an approximately exponential inflationary phase associated to $q^{(-)}(\frac3{16})=-1$, this solution is shown as a solid line. The other solution that possesses an early time $q^{(+)}(\frac3{16})=\frac12$-behavior with a Big Bang is shown as dotted line.
\label{fig:parameter_survey_first_solution_variations}
}
\end{minipage}
\end{figure}

As we see in Figure~\ref{fig:parameter_survey_first_solution_variations} (a), at $\xi=3/16$ we have a distinguished early time behavior. Namely, $q^{(-)}(3/16)=-1$, which coincides with the usual Dark Energy coefficient of a cosmological constant. So the $q^{(-)}(\xi)$-type expansion of the third solution class (III) has a Big Bang if $\frac16<\xi<\frac3{16}$ (i.e.\ $q^{(-)}(\xi)>-1$), whereas it corresponds to negative-power-expansion if $\xi>\frac3{16}$.  

At $\xi=\frac3{16}$, from our above point of view, we expect an exponential phase at early times. In fact, we can observe such a phase in the solution shown in Figure~\ref{fig:parameter_survey_first_solution_variations} (b) (with positive $\Lambda$ and $m$, but this effect shows up independently of $\Lambda$ and $m$). Recall that in Section~\ref{sec:Minkowski_and_De_Sitter_solutions} we observed the existence of a de Sitter solution for a massless field with precisely this $\xi$-value (even though for positive $\Lambda$ only). 

So our model includes an early inflationary phase without a large field mass or a self-interacting potential. The possibility of inflation with a non-minimally coupled scalar has been studied before \cite{Faraoni:1998qx}. In that reference, it was noted that we cannot have an inflationary phase for $\xi<1/6$ which we confirm (numerically) in the subsequent section. However, the solutions for $\xi>1/6$ were not studied, which is done in the present work.  

We forgo to study this inflationary effect any further and we note that a simple cross-check for the quality of a numerically obtained solution by evaluating $f$ along solutions (i.e.\ evaluating how well the energy equation holds) yields that the numerics is not very resilient after a few magnitudes of inflationary expansion. Thus it is unclear if this inflationary period lasts long enough to solve the cosmological problems such as the horizon problem or generate the density fluctuations observed in the CMB. In Figure~\ref{fig:parameter_survey_first_solution_variations} (b) we plot the part of the solution where we found the numerics to be trustworthy. The parameter setting $\xi=\frac3{16}$ of our model should rather be addressed with numerical DAE solvers.

\subsection{Small Bang solutions}
\label{sec:Small_Bang_solutions}

In this section we study the parameter regime of $0<\xi<\frac16$, and as before, we fix $\Lambda$ and $m$ and vary $q_0$.

We recall from Section~\ref{sec:Phase_space_constraints_from_the_energy_equation} that the solution set of the energy equation in $(\phi,\dot\phi,H)$-space is essentially (up to some spurious parameter settings) condensed in $-\phi_{\textup{crit}}(\xi)<\phi<\phi_{\textup{crit}}(\xi)$. Moreover, we have shown that the solution set comes arbitrarily close to the planes defined by $\pm\phi_{\textup{crit}}(\xi)$ at large $H$. On the other hand, the dynamical vector field \eqref{eqn:syst} is singular at these $\phi$-values. We will see that in fact, solutions generically approach this regime. Moreover, we will see that this regime is approached fast enough so that the resulting cosmology has a singularity even before the scale factor reaches zero and thus, we observe singularities without a spatially degenerating metric in the limit. 
We call such behavior Small Bang singularities as defined in Section~\ref{sec:Types_of_cosmological_singularities}.

In the case $0<\xi<\frac16$ both the determinant and the trace in \eqref{eq:det_and_trace_for_conic_section_matrix} are positive for all $H>0$, independently of $m$. Hence, the energy equation is solved by an ellipse in the $(\phi,\dot\phi)$-plane if $H>H_{\textup{vac}}(\Lambda)$, only by the point $(\phi,\dot\phi)=(0,0)$ if $H=H_{\textup{vac}}(\Lambda)$ and has no solutions in the $(\phi,\dot\phi)$-plane if $H<H_{\textup{vac}}(\Lambda)$. Consequently, any solution is strictly increasing or decreasing and, as before, we restrict ourselves to the increasing ones. Moreover, we recall that these ellipses become stable as $H\to\infty$ in the blow-up substitution $\dot\phi\mapsto H\dot\phi$ or, if we set $m=\Lambda=0$, even do not depend on $H$ at all in this blow-up.

\begin{figure}
\centering
    \includegraphics{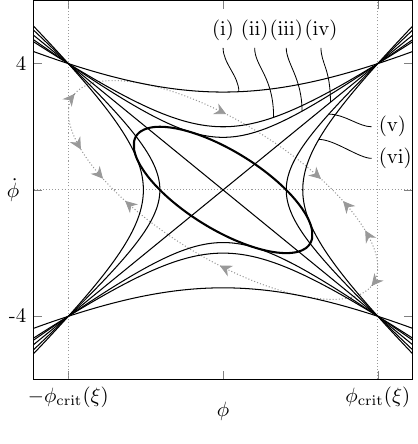}
    \qquad
\begin{minipage}[b]{.5\textwidth}
\captionsetup{format=plain, labelfont=bf,singlelinecheck=false}
\caption{
The graphic shows the conic sections solving the energy equation \eqref{eq:energy_equation_in_phase_space_section} as thick line and solving the trace equation \eqref{eq:trace_equation_with_q} as thin lines for the parameters $\xi=\frac{1}{12}$, $\Lambda=2$, $m=0$ and $H=1$. For the trace equation we chose the $q_0$-values:
\label{fig:parameter_survey_conic_sections_for_small_bangs}
}

\quad\includegraphics{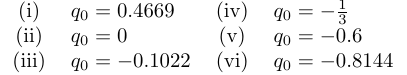}

\noindent
The gray dotted ellipse marks the limit conic section as $H\to\infty$ (equivalently, the $\Lambda=0$-conic section) if we read the plot to show conic sections in the $(\phi,H\dot\phi)$-plane. The arrows mark the direction of the time flow of the dynamics of our model in this limit.
\end{minipage}
\end{figure}

We pick the parameters $m=0$, $\Lambda=2$ and $\xi=\frac1{12}$ as generic for this regime. Figure~\ref{fig:parameter_survey_conic_sections_for_small_bangs} shows the conic sections solving the energy equation \eqref{eq:energy_equation_in_phase_space_section} (as the thick line ellipse) and the trace equation \eqref{eq:trace_equation_with_q} for a selection of $q_0$ values (as the thin line hyperbolas) at $H=1$ in the $(\phi,\dot\phi)$-plane. The $q_0$-values are chosen so that they contain some distinguished values and reflect the boundaries of admissible $q_0$ values. For cases (iii) and (vi) we have two choices of consistent initial values for the field which correspond to one cosmology. For cases (iv) and (v) we have four choices and two cosmologies. For any values of $q_0>-0.1022$ or $q_0<-0.8144$ no choices exist. This is visible in Figure~\ref{fig:parameter_survey_conic_sections_for_small_bangs} in that smaller or larger $q_0$ values, respectively, yield hyperbolas that lie outside of the ellipse. The critical value $\phi=\pm\phi_{\textup{crit}}(\xi)$ makes \eqref{eq:trace_equation_with_q} independent of $q$ and the solution set resembles the two straight lines (iv). Finally, Figure~\ref{fig:parameter_survey_conic_sections_for_small_bangs} includes as a gray dotted line the limit ellipse as $H\to\infty$ if we read the whole figure in the $(\phi,H\dot\phi)$-plane (which makes no difference at $H=1$). The arrows indicate the direction of the time flow of our model along this ellipse.

\begin{figure}
\begin{minipage}[b]{.5\textwidth}
    \includegraphics{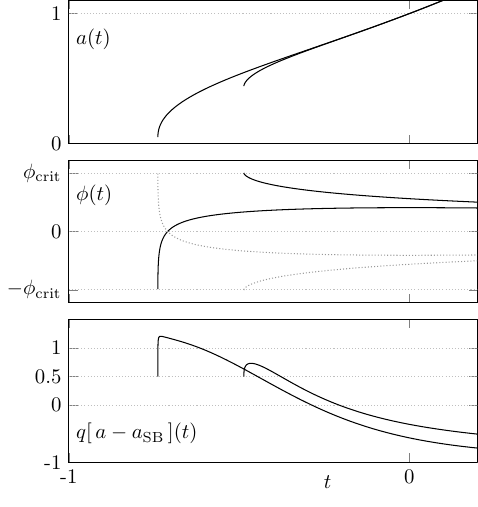}
\end{minipage}
\hfill
\begin{minipage}[b]{.45\textwidth}
\includegraphics{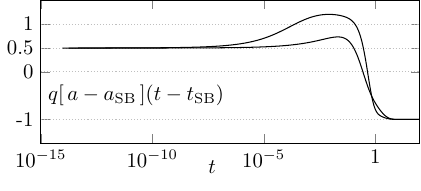}
\captionsetup{format=plain, labelfont=bf,singlelinecheck=false}
\caption{
The graphic shows the solutions to our model for the parameters $\xi=\frac{1}{12}$, $\Lambda=2$, $m=0$, $q_0=-0.6$ and $H=1$ in terms of $a$, $\phi$ and the deceleration parameter $q[\,a-a_{\textup{SB}}\,]$ where $a$ is shifted by the Small Bang value $a_{\textup{SB}}$. The right plot also shows $q[\,a-a_{\textup{SB}}\,]$ to the point where the numeric solver stopped, but with a logarithmic time axis shifted by the Small Bang time $t_{\textup{SB}}$.
\label{fig:parameter_survey_solutions_for_small_bangs}
}
\end{minipage}
\end{figure}

Figure~\ref{fig:parameter_survey_solutions_for_small_bangs} shows solutions with our present parameter choice in terms of the scale factor $a$ and the field $\phi$ (also with the equivalent field solutions as dotted lines). One can observe how for both solutions the field $\phi$ runs into the $\phi_{\textup{crit}}(\xi)$-singularity of the dynamic vector field \eqref{eqn:syst}. As we argued in Section~\ref{sec:Phase_space_constraints_from_the_energy_equation}, because of $\Lambda>0$ this necessarily entails that $H=\frac{\dot a}{a}\to\infty$ in that regime and, also recalling the monotonicity of $a$, the resulting spacetime necessarily has a singularity in the past.
Surprisingly, this happens fast enough so that this is not a Big Bang singularity, but a Small Bang at positive $a$ according to Definition~\ref{def:smallbang}. The time and scale factor of this singularity are labeled as $t_{\textup{SB}}$ and $a_{\textup{SB}}$, respectively.

It is interesting to study the deceleration parameter of these solutions. A plot of $q[\,a\,]$ shows a divergence $q[\,a\,](t)\to\infty$ as $t\to t_{\textup{SB}}$. However, if we shift the solution and study $q[\,a-a_{\textup{SB}}\,]$ instead, we observe (numerically) that this quantity does have a limit $q[\,a-a_{\textup{SB}}\,]\to\frac{1}{2}$ as $t\to t_{\textup{SB}}$. This is shown in Figure~\ref{fig:parameter_survey_solutions_for_small_bangs}, both in a linear and a logarithmic time-scale (the latter shifted by $t_{\textup{SB}}$). Notice that the plot with a logarithmic time-scale is very sensitive to the precise value of $a_{\textup{SB}}$ so a higher numerical precision was used. 

We recall that a $q$-value of $\frac{1}{2}$ or, equivalently by \eqref{eq:deceleration_state_fraction_correspondence}, a $\Gamma$-value of 0, is associated with a power-$\frac{2}{3}$-expansion. Hence, our numerical findings in Figure~\ref{fig:parameter_survey_solutions_for_small_bangs} on $q[\,a-a_{\textup{SB}}\,]$ suggest that 
\begin{equation}
\label{eq:expansion_of_scale_factor_around_aSB}
a(t)\in a_{\textup{SB}}+\alpha (t-t_{\textup{SB}})^{\nicefrac23}+\scalebox{.6}{$\mathbf{\mathcal{O}}$}\big((t-t_{\textup{SB}})^{\nicefrac23}\big)
\end{equation}
(small-\scalebox{.6}{$\mathbf{\mathcal{O}}$}) as $t\to t_{\textup{SB}}$, for a suitable $\alpha>0$. Such a solution has no analytic continuation, not even in a weak Puisseux sense. A Puisseux series is a Laurent series in $(t-t_0)^{\nicefrac1n}$ for some $n\in\mathbb N$ and $t_0\in\mathbb R$. By a Puisseux continuation we mean that for a scale factor $a$ on some interval $(t_0,t_0+\varepsilon)$ the graph $t\mapsto\big(t,a(t)\big)$ possesses an analytic continuation which can be represented as graph of a function of $t$ again. If so, this continuation can be determined by the Puisseux series of $a$ in $t_0$. In that case, $a$ may have a singularity in the sense that $\dot a$ (or some higher derivative) is unbounded, but still can be continuated continuously\footnote{For instance, the function $t\mapsto a_0+a_1(t-t_0)^{\nicefrac13}$ on $(t_0,t_0+\varepsilon)$ with some $a_0>0,a_1\neq0$ possesses a Puisseux continuation to values of $t$ smaller than $t_0$ while the function $t\mapsto a_0+a_1(t-t_0)^{\nicefrac23}$ does not.}.

\begin{figure}
\centering
    \includegraphics{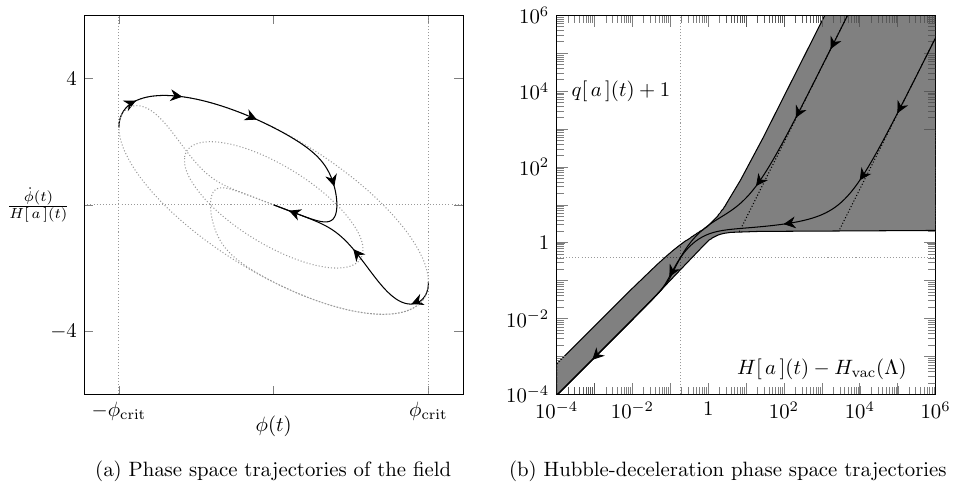}

\smallskip
\begin{minipage}[b]{.9\textwidth}
\captionsetup{format=plain, labelfont=bf,singlelinecheck=false}
\caption{
Plot (a) shows the trajectories of the field variables $\phi$ and $\dot\phi$ for two inequivalent solutions shown in Figure~\ref{fig:parameter_survey_solutions_for_small_bangs}. The other two solutions are shown as dotted gray trajectories. Hereby, $\dot\phi$ is normalized to show the trajectory in the $(\phi,H\dot\phi)$-plane. The arrows indicate the direction of time flow. The dotted ellipses mark the solution set to the energy equation at $H=1$ and $H\to\infty$, as in Figure~\ref{fig:parameter_survey_conic_sections_for_small_bangs}. Plot (b) shows the trajectories of these two solutions in the $H$-$q$-plane. Also here, the arrows indicate the time flow. The dotted black lines mark the asymptotics as described in the text. The gray area marks the points at which there exist consistent initial value pairs for $(\phi,\dot\phi)$, as described in Section~\ref{sec:deceleration_constraints}. Specifically, this color indicates the existence of four solutions.
\label{fig:parameter_survey_phase_space_trajectories_for_small_bangs}
}
\end{minipage}
\end{figure}

In Figure~\ref{fig:parameter_survey_phase_space_trajectories_for_small_bangs} (a), we show the trajectories of the curves
\begin{equation}
t\mapsto\bigg(\phi(t),\frac{\dot\phi(t)}{H[\,a\,](t)}\bigg)
\end{equation}
as obtained from each of the two solutions. The arrows indicate the time flow of our model and we see how these trajectories, including the time flow arrows, align with the limiting ellipse for $H \to \infty$.
Moreover, we observe (numerically) a convergence to the singularity points $\big(\pm\phi_{\textup{crit}}(\xi),\mp6\xi\phi_{\textup{crit}}(\xi)\big)$ in the Small Bang. From this convergence, together with the alignment to the limiting ellipse, we conclude that
\begin{equation}
\label{eq:limiting_energy_ellipse_around_pm_phicrit}
\frac{\dot\phi(t)}{H[\,a\,](t)}
\in
\mp\,6\xi\phi_{\textup{crit}}(\xi)
\mp \sqrt{\frac{12}{\phi_{\textup{crit}}(\xi)}} \sqrt{\phi_{\textup{crit}}(\xi) \mp \phi(t)}
+\mathcal{O}\big(\phi_{\textup{crit}}(\xi)\mp\phi(t)\big)
\end{equation}
as $\phi(t)\to\pm\phi_{\textup{crit}}$, by an expansion of the limiting ellipse (solving \eqref{eq:other_blowup_substituted_energy_conic_sections} for $\dot\phi$). 

In Figure~\ref{fig:parameter_survey_phase_space_trajectories_for_small_bangs} (b) we show the trajectories of the curve
\begin{equation}
t\mapsto\Big(H[\,a\,](t),q[\,a\,](t)\Big)\,,
\end{equation}
as obtained from each solution. This graphic also shows, as a gray area, the subset of $(H,q)$-pairs for which the energy equation possesses solutions for the field variables. One can see how in both the limits $t\to t_{\textup{SB}}$ and $t\to\infty$, the trajectories align to the boundary of the gray area. We note that the parallel straight lines in the loglog-plot indicate that the respective curves have a constant ratio. We find numerically with a high precision that these asymptotes at large $H$-values resemble quadratic functions in $H$ suggesting that
\begin{equation}
q[\,a\,](t)\in\alpha H[\,a\,](t)^2+\mathcal{O}\big(H[\,a\,](t)\big)
\end{equation}
as $t\to t_{\textup{SB}}$, for a suitable value $\alpha>0$.  By numerically evaluating $q[\,\phi_{\textup{crit}}(\xi)\pm\phi\,]$ (depending on whether $\phi(t)\to-\phi_{\textup{crit}(\xi)}$ or $\phi(t)\to\phi_{\textup{crit}(\xi)}$) we find the limit $q \to \frac{1}{2}$ as $t\to t_{\textup{SB}}$. This suggests that
\begin{equation}
    \phi(t)\in \pm\phi_{\textup{crit}}(\xi)\mp\beta (t-t_{\textup{SB}})^{\nicefrac23}+\scalebox{.6}{$\mathbf{\mathcal{O}}$}\big((t-t_{\textup{SB}})^{\nicefrac23}\big)
\end{equation}
in the same limit.

\begin{figure}[ht!]
\centering
    \includegraphics{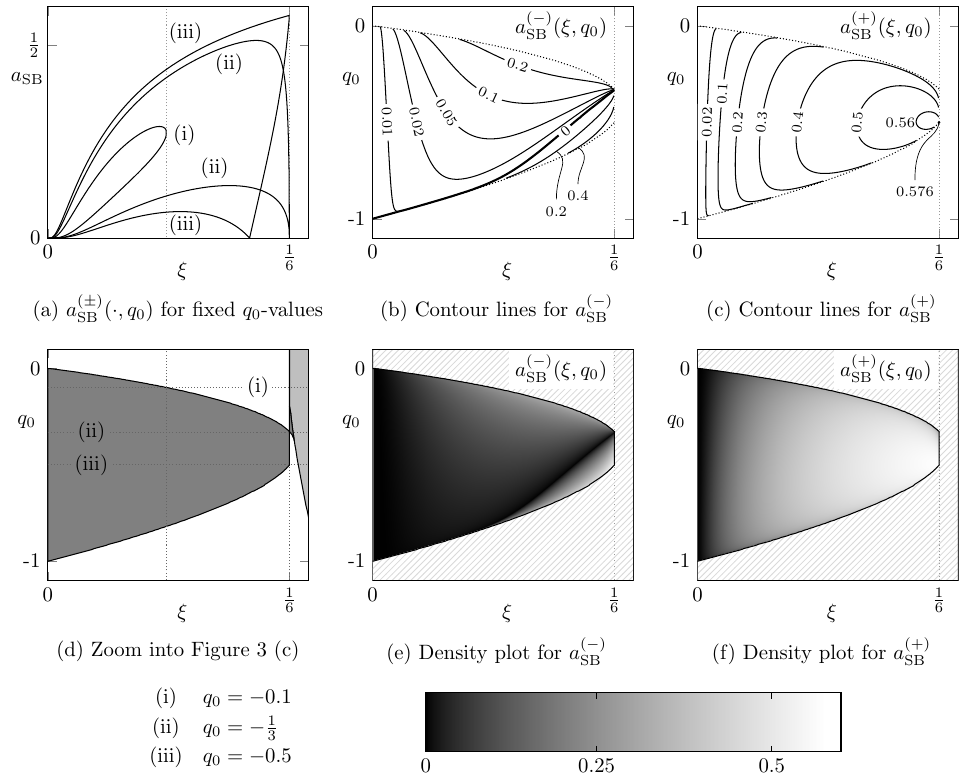}
\smallskip
\begin{minipage}[b]{.9\textwidth}
\captionsetup{format=plain, labelfont=bf,singlelinecheck=false}
\caption{
Plot (a) shows the graphs for $a_{\textup{SB}}^{(\pm)}(\cdot,q_0)$ for the $q_0$-values (i), (ii) and (iii). The gray dotted lines mark the upper bounds of the domains of these functions that are constrained by the existence of real initial values. Plot (d) shows a zoom into Figure~\ref{fig:initial} (c), where the domains of $a_{\textup{SB}}^{(\pm)}(\cdot,q_0)$ can be read off. Plots (b) and (c) show $a_{\textup{SB}}^{(\pm)}$ as functions of $\xi$ and $q_0$. Density Plots (e) and (f) also show the values of $a_{\textup{SB}}^{(\pm)}$.
\label{fig:parameter_survey_small_bangs_functions_of_xi_and_q}
}
\end{minipage}
\end{figure}

Our numerical results indicate that the leading term around $t_{\textup{SB}}$ is of order $(t-t_{\textup{SB}})^{2/3}$ \eqref{eq:expansion_of_scale_factor_around_aSB}. If we make the assumption that $a$ and $\phi$ actually possess a convergent Puisseux expansion in the Small Bang singularity of the form
\bml
\label{eq:assumption_on_expansions}
\begin{eqnarray}
\label{eq:assumption_on_expansion_of_a}
    a(t)&\in & a_{\textup{SB}}+\alpha (t-t_{\textup{SB}})^{\nicefrac23}+\mathcal{O}\big((t-t_{\textup{SB}})^{\nicefrac43}\big)
    \\
\label{eq:assumption_on_expansion_of_phi}
\phi(t)&\in & -\phi_{\textup{crit}}(\xi)+\beta (t-t_{\textup{SB}})^{\nicefrac23}+\mathcal{O}\big((t-t_{\textup{SB}})^{\nicefrac43}\big)
\end{eqnarray}
\eml
(w.l.o.g.\ we consider a solution with $\phi(t)\to-\phi_{\textup{crit}}(\xi))$ as $t\to t_{\textup{SB}}$) for suitable $\alpha,\beta>0$, then we find by term-wise differentiation that
\begin{equation}
\beta=\frac{\alpha}{a_{\textup{SB}}}6\xi\phi_{\textup{crit}}(\xi)
\end{equation}
is necessary for the limit $\dot\phi(t)/H[\,a\,](t)\to 6\xi\phi_{\textup{crit}}(\xi)$ (cf.\ \eqref{eq:limiting_energy_ellipse_around_pm_phicrit}). In fact, computing also $\ddot\phi$ and $\ddot a$, we find that the lowest (i.e. leading) order of $t-t_\textup{SB}$ occurring in any of the trace, energy or Klein-Gordon equation \eqref{eqn:KG_cosmology} cancel. We note that in view of quite recent developments in the theory of ODEs \cite{cano}, an assumption as in \eqref{eq:assumption_on_expansions} appears resilient. However, an applicability of these arguments needs to be carefully checked.

Finally, we discuss how the occurrence of Small Bangs is generic for $0<\xi<\frac16$.  Recall that we found in Section~\ref{sec:deceleration_constraints} that not all choices of parameters $0<\xi<\frac16$ and $q_0$ admit real initial values for $(\phi,\dot\phi)$. The region with real initial values is depicted in Figure~\ref{fig:initial} (c) with a zoom-in version in Figure~\ref{fig:parameter_survey_small_bangs_functions_of_xi_and_q} (d). Inside this region we found four pairs of initial values, hence our model admits up to two inequivalent cosmologies. Denote by
\be
a_{\textup{SB}}^{(+)}(\xi,q_0)
\qquad\textup{and}\qquad
a_{\textup{SB}}^{(-)}(\xi,q_0)
\ee
the larger and smaller value for $a_{\textup{SB}}$, respectively.  

Figure~\ref{fig:parameter_survey_small_bangs_functions_of_xi_and_q} (a) shows the graphs of the functions $\xi\mapsto a_{\textup{SB}}^{(\pm)}(\xi,q_0)$ for a selection of $q_0$-values, each on their maximal domains where consistent initial values for our model exist. We can see how towards the upper bound of the domain both $a_{\textup{SB}}^{(+)}(\cdot,q_0)$ and $a_{\textup{SB}}^{(\pm)}(\cdot,q_0)$ tend towards the same value and consequently, the union of both the graphs of both functions forms a closed curve. This is not a surprising behavior, recalling that the energy equation's conic section is an ellipse for $H[\,a\,](0)=1>H_{\textup{vac}}(\Lambda)$ at all $\xi\in(0,\frac16)$. Hence, if we continuously deform both the energy and the trace equation's conic section from a parameter choice with four intersection points to a parameter choice with an empty intersection, in between there must be a parameter point where the four intersection point degenerate into two intersection points. In this boundary case we thus observe one cosmology, particularly a tendency of $a_{\textup{SB}}^{(+)}(\xi,q_0)$ and $a_{\textup{SB}}^{(-)}(\xi,q_0)$ towards each other.

The most important feature of  Figure~\ref{fig:parameter_survey_small_bangs_functions_of_xi_and_q} is how generic the occurrence of a Small Bang is in the $0<\xi<1/6$ parameter regime. This is obvious from the plots of the level sets of $a_{\textup{SB}}^{(+)}$ and $a_{\textup{SB}}^{(-)}$ as functions of both $\xi$ and $q_0$ shown in Figure~\ref{fig:parameter_survey_small_bangs_functions_of_xi_and_q} (b) an (c) as well as the density plots (e) and (f). However, there is a regime of the parameter space where the Small Bangs degenerate into Big Bangs, $a_{\textup{SB}}^{(-)}\to0$, which is visible as the thick line marking the level-0-set in Figure~\ref{fig:parameter_survey_small_bangs_functions_of_xi_and_q} (b).

\subsection{Dark Energy dominated late times}
\label{sec:Dark_Energy_dominated_late_times}

We state that in every solution we studied with $\xi\ge0$ and $\Lambda>0$ we observed a de Sitter late-time expansion with $q[\,a\,](t)\to-1$ and $H[\,a\,](t)\to H_{\textup{vac}}(\Lambda)$ as $t\to\infty$.
We omit presenting a numerical proof, as that would simply show ``true'' at every parameter point. The fields asymptotically behave as $\phi(t)\to0$ and $\dot\phi(t)\to0$ in the same limit. 
Examples of such solutions are shown in Figures~\ref{fig:parameter_survey_first_solution_variations},~\ref{fig:parameter_survey_solutions_for_small_bangs} and \ref{fig:parameter_survey_phase_space_trajectories_for_small_bangs}. 

This effect is strongly reminiscent of the so-called cosmic no-hair theorem. This was proven classically by Wald \cite{Wald:1983ky}. In this work, it was shown that initially expanding cosmological models (particularly with our flat spatial sections) satisfying Einstein's equation with a positive cosmological constant evolve towards a de Sitter solution. The proof assumes two pointwise energy conditions, the weak and the strong
\begin{equation}
    T_\munu\, t^\mu t^\nu\ge0
    \qquad\textup{and}\qquad
    \Big(T_\munu-\frac12\, T^\sigma{}_{\!\sigma} g_\munu\Big)\,t^\mu t^\nu\ge0 \,,
\end{equation}
for all future-directed time-like vector fields $t^\mu$. Here, the physical and geometric forms of the conditions coincide as the cosmological models obey Einstein's equation. As we mentioned in Section~\ref{sec:Types_of_cosmological_singularities}, both of those conditions are violated by non-minimally coupled fields \footnote{The weak energy condition is obeyed for $\xi \neq 0$ while the strong energy condition is violated for any massive scalar or for $\Lambda >0$.}. However, averaged forms of both conditions exist \cite{Fewster:2006ti, Brown:2018hym} for non-minimal coupling. While there is no cosmic no-hair theorem for averaged conditions, our numerical results indicate the plausibility of such a theorem. 

Next, we want to discuss how the de Sitter point $\big(0,0,H_{\textup{vac}}(\Lambda)\big)$ in $(\phi,\dot\phi,H)$-space is approached. Recall from Section~\ref{sec:Minkowski_and_De_Sitter_solutions} that the solutions of the Klein-Gordon equation on a given de Sitter background were classified into the parameter regimes of a underdamped, critically damped or overdamped harmonic oscillator. 

We relax the the condition of an exact de Sitter background and study the approximate de Sitter late-time behavior of the Klein-Gordon field in our model. We can still decide whether the de Sitter point $\big(0,0,H_{\textup{vac}}(\Lambda)\big)$ lies in an under-, critically or overdamped regime. Formally, we distinguish whether the expression
\begin{equation}
\label{eq:dampedness_condition_in_late_time_section}
    M^2-\frac94H^2=m^2+12\Big(\xi-\frac{3}{16}\Big)H_{\textup{vac}}(\Lambda)^2
\end{equation}
is positive for the underdamped case, zero for the critically damped case and negative for the overdamped case. Notice how the de Sitter point is always in the underdamped regime for $\xi>\frac{3}{16}$. Moreover, we recall that for $\xi\in(0,\frac16)$ the energy equation \eqref{eq:energy_equation_in_phase_space_section} has no solution with $H<H_{\textup{vac}}(\Lambda)$. Hence, for these $\xi$-values, if the de Sitter point is overdamped (``$<$" in \eqref{eq:dampedness_condition_in_late_time_section}), any point solving the energy equation lies in the overdamped regime. Also for these $\xi$-values, if the de Sitter point is critically damped, any other point lies in an overdamped regime.

\begin{figure}
\centering
    \includegraphics{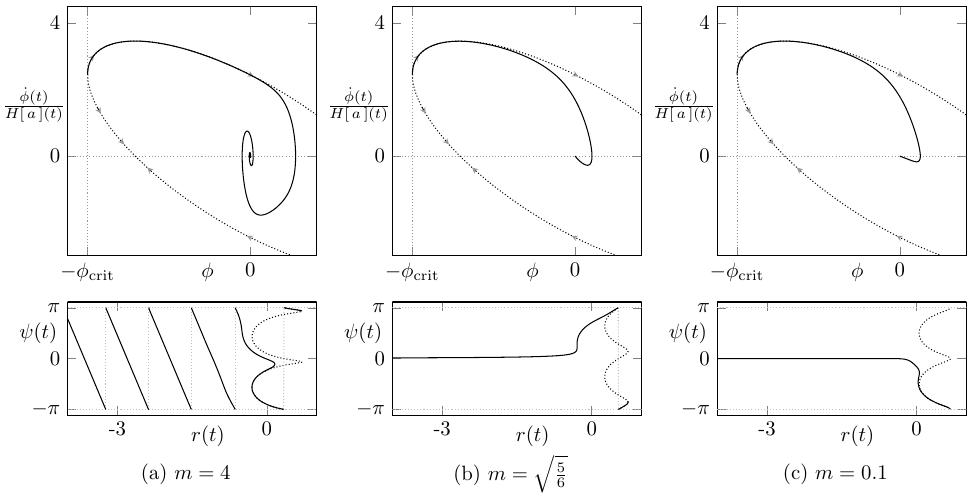}

\smallskip
\begin{minipage}[b]{.9\textwidth}
\captionsetup{format=plain, labelfont=bf,singlelinecheck=false}
\caption{
The graphics shows the trajectories of $t\mapsto\big(\phi(t),\frac{\dot\phi(t)}{H[\,a\,](t)}\big)$ for the parameters $a(0)=\dot a(0)=1$, $\Lambda=2$, $\xi=\frac1{12}$, $q_0=-0.2$ and masses as indicated in the subcaption's titles. Each case is presented both in Cartesian (top) as well as in polar (bottom) coordinates $r$ and $\psi$, where for the latter we used a logarithmic scaling for the radial coordinate and we applied a certain linear coordinate transformation as described in the text. The dotted lines show the limiting ellipses for the energy equation as $H\to\infty$ that align each trajectory at early times.
\label{fig:late_time_regimes}
}
\end{minipage}
\end{figure}

We want to study a similar parameter setting as in the previous section, that is, $\Lambda=2$ and $\xi=\frac1{12}$ (which is $<\frac3{16}$), but we choose positive masses $m$ in order to cover all cases of dampening. 
Notice that $m=\sqrt{\nicefrac56}$ is the critical value imposing that the expression in \eqref{eq:dampedness_condition_in_late_time_section} vanishes for our chosen $\Lambda$ and $\xi$. A larger or smaller mass means a underdamped or overdamped de Sitter point, respectively, and we pick $m=4$ and $m=0.1$ as representative for these two regimes. We have four initial value pairs $(\phi,\dot\phi)$ for each chosen $m$-value if we choose $q_0=-0.2$. As usual, we stick with $a(0)=\dot a(0)=1$. Figure~\ref{fig:late_time_regimes} shows the trajectories for
\begin{equation}
\label{eq:trajectory_in_late-time_section}
t\mapsto\bigg(\phi(t),\frac{\dot\phi(t)}{H[\,a\,](t)}\bigg)
\end{equation}
for our parameter choice for the tree different masses (upper plots). 
We chose in each regime one initial value pair for $(\phi,\dot\phi)$, however, we remark that the other solutions (particularly, the ones leading to an inequivalent cosmology) exhibit similar behavior.

The lower graphics in Figure~\ref{fig:late_time_regimes} are for the same parameters as the respective upper plots, but for the following trajectories
\begin{equation}
    t\mapsto
    \begin{pmatrix}
        r(t)\\\psi(t)
    \end{pmatrix}
    :=
    \begin{pmatrix}
        \frac{1}{\log(10)}&0\\0&1
    \end{pmatrix}
    \log\left(C
    \begin{pmatrix}
        1&0\\0&\frac{H_{\textup{vac}}(\Lambda)}{H[\,a\,](t)}
    \end{pmatrix}
    \begin{pmatrix}
        \phi(t)\\\dot\phi(t)
    \end{pmatrix}
    \right)\,.
\end{equation}
The first transformation (the rightmost matrix) multiplies $\dot\phi$ by $H[\,a\,]$ in accordance with the respective upper plots of the trajectories \eqref{eq:trajectory_in_late-time_section}. Moreover, it is normalized by the factor $H_{\textup{vac}}(\Lambda)$ so we only distort the trajectories away from the de Sitter point $(\phi,\dot\phi,H)=\big(0,0,H_{\textup{vac}}(\Lambda)\big)$. The logarithm is understood as the map $\log:\mathbb R^2\to\mathbb R^2$ given by the principal branch of the usual logarithm, identifying $\mathbb R^2\cong\mathbb C$. Equivalently, the logarithm maps to polar coordinates with a logarithmic radial component. The choice of the principal branch imposes an imaginary part/angle coordinate in $(-\frac\pi2,\frac\pi2)$. The left-most transformation adjusts the logarithmic radius into a base-10-logarithm. Finally, the matrix $C$ is a linear transformation depending on whether the de Sitter point lies in an under-, critically or overdamped regime and should gain comparability of our solutions here with the solutions of the Klein-Gordon equation on an exact de Sitter background in Section~\ref{sec:Minkowski_and_De_Sitter_solutions}. They are given as follows: 

In the underdamped case $m^2+12\Big(\xi-\frac{3}{16}\Big)H_{\textup{vac}}(\Lambda)^2>0$ one can compute for a underdamped solution on a de Sitter background 
\begin{equation}
    \begin{pmatrix}
1&0
\\
-\frac32H
&-\tfrac12\sqrt{4M^2-9H^2}
\end{pmatrix}^{-1}\begin{pmatrix}
\phi_{\mathrm{ud}}(t)
\\
\dot\phi_{\mathrm{ud}}(t)
\end{pmatrix}
=
\mathrm{e}^{-\frac32Ht}\begin{pmatrix}
\cos\Big(\tfrac12\sqrt{4M^2-9H^2}\,t\Big)
\\[5pt]
\sin\Big(\tfrac12\sqrt{4M^2-9H^2}\,t\Big)
\end{pmatrix}\,,
\end{equation}
where $\phi_{\textup{ud}}$ and $M^2$ are given in Section~\ref{sec:Minkowski_and_De_Sitter_solutions}. Thus, after the transformation by 
\begin{equation}
    C_{\textup{ud}}(H):=\begin{pmatrix}
1&0
\\
-\frac32H
&-\tfrac12\sqrt{4M^2-9H^2}
\end{pmatrix}^{-1}
\end{equation}
the phase space trajectory of $\phi_{\textup{ud}}$ is a logarithmic spiral. Hence, we expect that after transformation with $C_{\textup{ud}}\big(H_{\textup{vac}}(\Lambda)\big)$ also our trajectories are well approximated by such a logarithmic spiral. Indeed, we can observe in Figure~\ref{fig:late_time_regimes} (a) that after being aligned with the large-$H$-limit ellipse (the dotted black line on the right, using the same transformations), the trajectory goes over into a clean saw-tooth function. Notice that a logarithmic spiral in polar coordinates with logarithmic radii is precisely such a saw-tooth function.

In the critically damped regime $m^2+12\Big(\xi-\frac{3}{16}\Big)H_{\textup{vac}}(\Lambda)^2=0$ we compute 
\begin{equation}
\label{eq:crit_damped_late_time}
\begin{pmatrix}
\phi_{\mathrm{cd}}(t)
\\
\dot \phi_{\mathrm{cd}}(t)
\end{pmatrix}
=
c_{1}\,\mathrm{e}^{-\frac32Ht}\begin{pmatrix}
1
\\
-\frac32H
\end{pmatrix}
\qquad\textup{or}\qquad
\begin{pmatrix}
\phi_{\mathrm{cd}}(t)
\\
\dot \phi_{\mathrm{cd}}(t)
\end{pmatrix}
\in
c_{2}\,t\,\mathrm{e}^{-\frac32Ht}\begin{pmatrix}
1
\\
-\frac32H
\end{pmatrix}
+\mathcal{O}\big(\mathrm{e}^{-\frac32Ht}\big)
\end{equation}
as $t\to\infty$ (component-wise big-$\mathcal{O}$) for $c_2=0$ or $c_2\neq0$, respectively, and we discard the trivial case $c_1=c_2=0$. Hence, we expect our solution running into a critically damped de Sitter point to be aligned with the direction of the vector in \eqref{eq:crit_damped_late_time} and we set 
\begin{equation}
    C_{\textup{cd}}(H):=
    \frac{1}{\sqrt{1+\frac94H^2}\,}
    \begin{pmatrix}
        1&-\frac32H\\\frac32H&1
    \end{pmatrix}\,,
\end{equation}
the rotation matrix with angle $\arctan(-\frac32H)$. In fact, our trajectories after the transformation $C_{\textup{cd}}\big(H_{\textup{vac}}(\Lambda)\big)$ in Figure~\ref{fig:late_time_regimes} (b) show an alignment with the $\big(\!\begin{smallmatrix}1\\0\end{smallmatrix}\!\big)$-direction as $\psi(t)\to0$ and $r(t)\to-\infty$.

Finally, in the overdamped case $m^2+12\Big(\xi-\frac{3}{16}\Big)H_{\textup{vac}}(\Lambda)^2<0$, we compute 
\begin{equation}
\label{eq:overdamped_late_time}
    \begin{pmatrix}
1&1
\\
\lambda_+&\lambda_-
\end{pmatrix}^{-1}
\begin{pmatrix}
\phi_{\mathrm{od}}(t)
\\
\dot\phi_{\mathrm{od}}(t)
\end{pmatrix}
=
\begin{pmatrix}
c_1\mathrm{e}^{\lambda_+t}
\\
c_2\mathrm{e}^{\lambda_-t}
\end{pmatrix}
\\
\in c_1\mathrm{e}^{\lambda_+t}\begin{pmatrix}1\\0\end{pmatrix}+\mathcal{O}\big(\mathrm{e}^{\lambda_-t}\big)
\end{equation}
as $t\to\infty$, with the roots $\lambda_\pm$ of the Klein-Gordon equations characteristic polynomial. We read off from \eqref{eq:charpol_roots_KG_on_dS} that generally $\lambda_+>\lambda_-$ and we note that for our presently chosen parameters $\lambda_+<0$.
Hence, we set
\begin{equation}
    C_{\textup{od}}(H):=
    \begin{pmatrix}
1&1
\\
\lambda_+&\lambda_-
\end{pmatrix}^{-1}
\end{equation}
and expect that after transformation with $C_{\textup{od}}\big(H_{\textup{vac}}(\Lambda)\big)$ our trajectory is aligned with  the $\big(\!\begin{smallmatrix}1\\0\end{smallmatrix}\!\big)$-direction. Also in this case, Figure~\ref{fig:late_time_regimes} (c) shows such an alignment as $\psi(t)\to 0$ and $r(t)\to-\infty$.

While the underdamped case (a) has a very different late-time behavior than the other two cases, these are in turn quite similar. The main difference between (b) and (c) lies in how the de Sitter point is approached. The trajectory in (c) appears to straighten up before eventually approaching its limit, while the trajectory in (b) does not show this behavior. In fact, comparing the convergence rates in the expansions in \eqref{eq:crit_damped_late_time} (the $c_2\neq 0$-case) and \eqref{eq:overdamped_late_time} one would expect the trajectory in (c) to align its ``limit direction" much faster. In fact,  the polar plots appear as if in (c) the limit $\psi(t)\to0$ is approached faster than in (b).

\subsection[Example with \texorpdfstring{$\Lambda$}{sth}CDM values]{Example with \texorpdfstring{$\mathbf\Lambda$}{sth}CDM values}

We conclude the discussion of the behavior of the scale factor in our model with an example of a massless field. The parameters $\Lambda$, $H_0$ and $q_0$ are taken from the $\Lambda$CDM model, whereas $\xi$ is used to tune the model. As we will see, this example exhibits all three types of early time behavior for different values of the coupling constant $\xi$, namely, a Big Bang, a Small Bang and the early time inflation phase.

For the observed value of the Hubble constant we use $H_0=2.19 \times 10^{-18} \, \text{s}^{-1}$ using results \cite{Planck-Collab} from the PLANCK collaboration\footnote{We are aware of the of the observational inconsistency of the value of the Hubble constant between observations of the CMB and from local measurements of distances and redshifts \cite{DiValentino:2021izs}. This difference however should not change the qualitative nature of our results in which the Hubble constant merely fixes a time scale}. The initial deceleration parameter is set to the $\Lambda$CDM value $q_0=-0.538$. We set the cosmological constant using the observed $\Omega_\Lambda$ value, which gives $\Lambda=9.95 \times 10^{-36}\,\text{s}^{-2}$.

\begin{figure}
    \centering
    \includegraphics{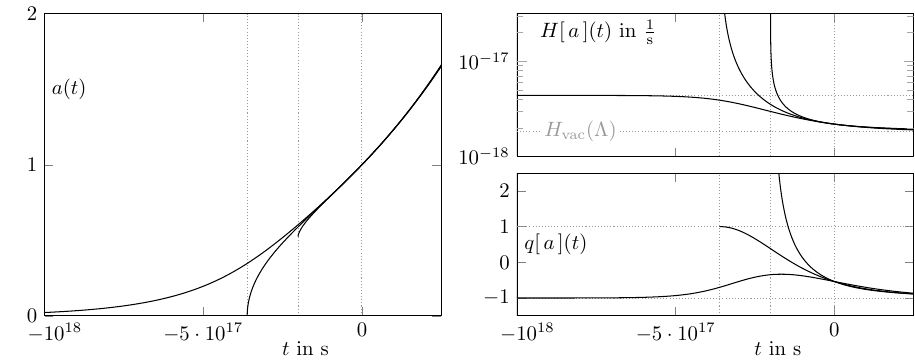}
    \begin{minipage}[b]{.9\textwidth}
    \captionsetup{format=plain, labelfont=bf,singlelinecheck=false}
    \caption{The graphics show three solutions using $m=0$ and the $\Lambda$CDM values for $\Lambda$, $q_0$ and $H_0$ as in the text, in terms of $a$, $H[\,a\,]$ and $q[\,a\,]$. $\xi$ is chosen as follows: The two solutions showing a Small Bang or a Big Bang are the two inequivalent solutions for $\xi=0.14215272$, whereas the other solution is for $\xi=\frac3{16}$. The dotted vertical lines mark distinguished times, today, the time of the Small Bang and of the Big Bang. The dotted horizontal lines mark distinguished values in the respective plot. We marked $H_{\textup{vac}}(\Lambda)$ which is approached by all three solutions at late times and the inflationary $H$ which is approached by the singularity-free ($\xi=\frac3{16}$) solution. Moreover, we marked the Dark Energy and the Radiation values $q=\pm1$.\label{fig:example}}
    \end{minipage}
\end{figure}

As we see in Figure~\ref{fig:example} these parameters exhibit all discussed solutions by choosing different values of $\xi$. First, setting $\xi=\frac3{16}$ gives an early inflationary phase for all possible initial values (see Section~\ref{sec:inflation}). As the allowed values for that choice of parameters all lie between $0<\xi<1/6$ or for $\xi>1/6$ we observe the following behaviors: for $\xi>1/6$ we have only two solutions (one cosmology) where the scale factor asymptotically approaches zero as $t\to -\infty$. Within that range we have the inflation value of $\xi=\frac3{16}$. Within the $0<\xi<1/6$ range we have all four solutions (two inequivalent cosmologies). As discussed in Section~\ref{sec:Small_Bang_solutions} this range of $\xi$ generically gives Small Bang solutions. However, for one of the two cosmologies there exists a value of $\xi$ for each $q_0\in(-1,-\frac13)$ where the Small Bang degenerates into a Big Bang, that is, $a(t)\to0$ (see Figure~\ref{fig:parameter_survey_small_bangs_functions_of_xi_and_q} (b)). For the $\Lambda$CDM value of $q_0$ this is $\xi=0.1421$. We see that the time of the Big Bang singularity is a bit less than the estimated age of the universe $-3.615 \times 10^{17}$s. The other cosmology for $\xi=0.1421$ is giving a Small Bang, where the scale factor is $0.525$ when its derivatives diverge. 

We should note that all three solutions (as all solutions with $\Lambda >0$ observed) exhibit a late time Dark Energy behavior. All three solutions have $q[a](t) \to -1$ and $H[a](t) \to H_{\text{vac}}$ at late times. 

It is interesting to observe that we do not find any bounce solutions. The bounce models in the literature are either semiclassical \cite{Parker:1973qd} or in the context of quantum gravity \cite{BarcaGiovannettiMontani:2021}. If they are classical, they employ unknown matter fields or alternative theories of gravity \cite{Battefeld:2014uga}.

\section{Summary and discussion}
\label{sec:discussion}

In this work, we performed an almost complete analysis of classical Klein-Gordon cosmology. While our scalar field was not interacting, we allowed for all values of mass, cosmological constant and coupling to curvature. For early times, we observed Big Bang solutions and Small Bang solutions, where the latter are quite generic in this model for a curvature coupling $0<\xi<\frac16$. We also observed asymptotic early time solutions entering inflationary era, particularly at $\xi=\frac3{16}$. This is a particularly interesting feature of our model, as inflation is achieved without any further assumptions or a self-interacting potential (resembling an effective cosmological constant). We did not observe any bounce solutions. At late times, all solutions with a positive cosmological constant are Dark Energy solutions pointing to a more general cosmic no-hair theorem than Wald's version.

One direction to be pursued in the future is whether a self-interaction of the field can be employed to weaken or even fully eradicate the effect of Small Bangs. In potentials like $\frac{\lambda}{2}\phi^4$ or $\lambda\cosh(\mu\phi)$ (``sinh-Gordon'' field), or even $\lambda/(\phi^2-\phi_\textup{crit}(\xi)^2)$, the field is expected to be stronger confined towards $0$, and thus away from $\phi_\textup{crit}(\xi)$. A purely classical theory for such potentials is straight-forward, although the conic sections will be replaced by more complicated algebraic or analytic varieties. In the context of a student project supervised by one of us, indications were found that a $\frac{\lambda}{2}\phi^4$ potential is indeed capable of degenerating the Small Bang of a comparable free field into a Big Bang, similarly to the effect shown in Figure~\ref{fig:parameter_survey_small_bangs_functions_of_xi_and_q} but with $a_\textup{SB}$ dependent on $\lambda$.

Finally, the motivation for this work, and the obvious extension, is to apply these results to analyze a semiclassical Klein-Gordon cosmological model. In the present article we put a special emphasis on how, to a certain extend, the dynamics of classical Klein-Gordon can be understood in terms of conic sections in phase space. In particular we plan to use the solutions as the one-point functions of the semiclassical stress-energy tensor \eqref{eq:SCE_with_split_QSE_tensor}. Notice that the principal structure of the energy equation and the trace equations as conic section equations, where the defining parameters are functions of the remaining phase space variables (in that model $a,\dot a,\ddot a$ and $a^{(3)})$, remains unchanged. It would be interesting to see which features of the classical cosmology remain present in the semiclassical case, with a focus on the behavior of solutions running into (or asymptotically towards) regions of phase space that are forbidden by the algebraic constraints. In a preliminary numerical survey we found that this is not the case, leading to the conjecture that quantum effects are capable of ``repairing" the unphysical behavior of Small Bangs. Also, a focus should be put on the influence of the de Sitter phase at $\xi=\frac{3}{16}$ and the ``quantum branch de Sitter solutions" from \cite{bd_on_ds} on each other. A third point would be to compare the inflationary early phases of the present work (at $\xi=\frac{3}{16}$) with similar solution found in \cite{Gottschalk:2021pkr} (at $\xi=\frac16$) as a solution of the cosmic horizon problem. However, a more systematic investigation is left for the future.

\section*{Acknowledgments}
The authors would like to thank Paul Anderson, Chris Fewster, Hanno Gottschalk, Daan Janssen, Joseph Nyhan, Ken Olum, Lennart Papp and Daniel Siemssen for useful conversations at different stages of the project. The authors acknowledge financial support from the London Mathematical Society (LMS). The ``Research in pairs'' LMS grant financed NR's visit to King's College London where part of this work was performed.

\newpage

\bibliographystyle{utphys}
\bibliography{references.bib}

\end{document}